%% file: Carpenter.tex
\documentclass[preprint]{aastex}

\received{}
\accepted{}
\journalid{}{}
\articleid{}{}

\newcommand{\ts}{\thinspace}
\newcommand{\simless}{\mathbin{\lower 3pt\hbox
     {$\rlap{\raise 5pt\hbox{$\char'074$}}\mathchar"7218$}}}
\newcommand{\simgreat}{\mathbin{\lower 3pt\hbox
     {$\rlap{\raise 5pt\hbox{$\char'076$}}\mathchar"7218$}}}

\newcommand{\about}    {$\sim$\ts}
\newcommand{\aboutless}{$\simless$\ts}
\newcommand{\aboutmore}{$\simgreat$\ts}

\newcommand{\msun}{\ts M$_\odot$}
\newcommand{\cmg}{\ts cm$^2$\ts g$^{-1}$}
\newcommand{\mjybeam}{\ts mJy\ts beam$^{-1}$}

\newcommand{\etal}{et~al.}

\begin{document}

\title{Evolution of Cold Circumstellar Dust Around Solar-Type Stars}

\author{John M. Carpenter\altaffilmark{1}}
\email{jmc@astro.caltech.edu}

\author{Sebastian Wolf\altaffilmark{1,2}}
\email{swolf@mpia.de}

\author{Katharina Schreyer\altaffilmark{3}}
\email{martin@astro.uni-jena.de}

\author{Ralf Launhardt\altaffilmark{2}}
\email{rl@mpia.de}

\and

\author{Th. Henning\altaffilmark{2}}
\email{henning@mpia.de}

\altaffiltext{1}{California Institute of Technology, Department of Astronomy, MS 105-24, Pasadena, CA 91125}
\altaffiltext{2}{Max Planck Institut f\"ur Astronomie, K\"onigstuhl 17, D-69117,
   Heidelberg, Germany}
\altaffiltext{3}{Astrophysikalisches Institut und Universit\"ats-Sternwarte, 
   Schillerg\"asschen 2-3, D-07745 Jena, Germany}

\begin{abstract}

We present submillimeter (CSO 350\micron) and millimeter (SEST 1.2~mm,
OVRO 3~mm) photometry for 125 solar-type stars from the FEPS {\it Spitzer} 
Legacy program that have masses between \about 0.5 and 2.0\msun\ and ages from 
\about 3~Myr to 3~Gyr. Continuum emission was 
detected toward four stars with a signal to noise ratio $\ge 3$: the classical 
T~Tauri stars RX~J1842.9$-$3532, RX~J1852.3$-$3700, and PDS~66 with SEST, and 
the debris disk system HD~107146 with OVRO. RX~J1842.9$-$3532 and 
RX~J1852.3$-$3700 are located in projection nearby the R~CrA molecular 
cloud with estimated ages of \about 10~Myr \citep{Neuhauser00}, while 
PDS~66 is a probable member of the \about 20~Myr old Lower Centaurus-Crux 
subgroup of the Scorpius-Centaurus OB association \citep{Mamajek04}. 
The continuum emission toward these three 
sources is unresolved at the 24$''$ SEST resolution and likely originates from 
circumstellar accretion disks, each with estimated dust masses of 
\about 5$\times10^{-5}$\msun. Analysis of the visibility data toward 
HD~107146 (age \about 80-200~Myr) indicates that the 3~mm continuum emission 
is centered on the star within the astrometric uncertainties and resolved with 
a gaussian-fit FWHM size of $(6.5''\pm1.4'') \times (4.2''\pm1.3'')$, 
or 185~AU$\times$120~AU. The results from our continuum survey are combined 
with published observations to quantify the evolution of dust mass with time
by comparing the mass distributions for samples with different stellar
ages. The frequency distribution of circumstellar dust masses around 
solar-type stars in the Taurus molecular cloud (age \about 2~Myr) is 
distinguished from that around 3-10~Myr and 10-30~Myr old stars at a 
significance level of \about1.5$\sigma$ and \about 3$\sigma$ respectively.
These results suggest a decrease in the mass of dust contained in small
dust grains and/or changes in the grain properties by stellar ages of 
10-30~Myr, consistent with previous conclusions. Further observations are 
needed to determine if the evolution in the amount of cold dust occurs on 
even shorter time scales.

\end{abstract}


\section{Introduction}

Most young (\about 1~Myr) solar mass stars are surrounded by circumstellar 
disks \citep{Strom89}. These disks have masses \citep{Beckwith90} and sizes 
\citep{McCaughrean96,Dutrey96} comparable to that expected for the primitive 
solar nebula, and are thus the suspected sites for the formation of planetary 
systems. Indeed, the increasing number of planets discovered around 
main-sequence stars \citep{Marcy00} strongly supports the notion 
that planet formation in circumstellar disks is a common occurrence. The 
physics of how planets form within circumstellar disks is unclear though
and remains the subject of ongoing investigation (cf. Pollack \etal~1996 and 
Boss~2001). Since the direct detection of planets around young stars is 
challenging for current instrumentation, observational constraints on the 
planet formation process are frequently inferred from the evolutionary time 
scales of circumstellar disks.

The evolution of circumstellar disks has been studied primarily in the
infrared by identifying stars that exhibit emission in excess of that expected
from the stellar photosphere. Near-infrared $JHKL$ band photometry of nearby 
star-forming regions \citep{Strom89,Haisch01}, which probes the inner 
\about 0.1~AU of the disk around solar-type stars, suggest that the percentage 
of stars with 2-4\micron\ excesses is \aboutmore 80\% for \about 1~Myr stars 
and diminishes to \about 50\% by an age of \about 3~Myr. The inner disk does
persist to relatively old ages in at least some stars, as evidenced by the 
high fraction of stars (60\%) with $L$-band excesses in the 5-9~Myr 
$\eta$~Cha cluster \citep{Lyo03}, the $K$-band excess in the 7-17~Myr star 
PDS~66 \citep{Mamajek02,Mamajek04}, and the disk accretion signatures in at
least two members of the \about 10~Myr old TW~Hydra association 
\citep{Muzerolle00}. By ages of \aboutless 10-15~Myr, 
the inner disk as traced by $K$-band photometric excesses has 
diminished to nearly zero (Mamajek, Meyer, \& Liebert~2002).
Similarly, 10\micron\ observations 
show that \about 20\% of the stars in the TW Hydra association 
\citep{Jay99,Wein04} exhibit evidence for dust between 0.1~AU and 1~AU,
with the excess fraction falling to approximately 0 by 30~Myr in at least the 
Tucana-Horologium association \citep{Mamajek04}. Far-infrared observations
obtained with IRAS and ISO suggest disk evolution from 1-10~AU on similar
time scales \citep{Meyer00}, but are inconclusive on whether subsequent
evolution is continuous \citep{Spangler01} or discrete 
\citep{Habing99,Habing01,Decin03}.

Millimeter-wavelength continuum observations provide a complimentary picture 
of disk evolution by probing colder disk material that is not 
detectable in the near- and mid-infrared. Continuum surveys of stars
in the Taurus (Beckwith \etal~1990, Osterloh \& Beckwith~1995; see also 
Motte \& Andr\'e~2001), $\rho$~Oph (Andr\'e \& Montmerle 1994; see also 
Motte, Andr\'e, \& Neri~1998, N\"urnberger \etal~1998, Johnstone \etal~2000),
and Chamaeleon I\&II \citep{Henning93} molecular clouds suggest that the 
median disk mass (including gas and dust components) around low-mass stars 
is \about 0.005\msun\ for stellar ages of \about 1~Myr. Moreover,
\citet{Beckwith90} found no evidence of temporal evolution in the mass of 
cold, small (\aboutless 1~mm) dust particles between ages of 0.1 and 10~Myr. 
\citet{Zuckerman93} conducted a continuum survey of stars in the Pleiades and 
Ursa Major and found that the dust masses decrease by at least two orders of 
magnitudes by an age of \about 300~Myr. Similar surveys of Lindroos binaries 
(A- and B- stars with ages of \about 10-200~Myr; see Jewitt~1994, Gahm 
\etal~1994, Wyatt, Dent, \& Greaves~2004) and a limited sample of stars in 
the $\beta$~Pic Moving Group and Local Association \citep{Liu04} suggest that 
few massive disks (\aboutmore 0.01\msun) survive beyond an age of 
\about 10~Myr.

Despite the substantial progress made by recent observations, our 
empirical understanding of the evolution of cold circumstellar dust remains 
incomplete. Analysis of the stellar ages in the Taurus molecular cloud 
(from which the Beckwith \etal~1990 sample was drawn) suggest that the 
apparent age spread can be attributed predominantly to observational
uncertainties \citep{Hartmann01}, and, therefore, studies of Taurus alone are 
insufficient to establish the time scales for disk evolution. The youngest
nearby ($<$ 200~pc) open clusters (e.g. Pleiades, IC~2602, $\alpha$~Per)
are frequently used to characterize the disk properties at ages of 
50-120~Myr. Few stars have been studied at intermediate ages (3-30~Myr)
though when planets of all masses are thought to be in their final assemblage 
stages. Lindroos binaries provide a few examples of these intermediate-age 
stars, but since these are systems where by definition the primary is an 
intermediate-mass star, they may not share the evolutionary history of 
``typical'' solar-type stars.

\begin{figure}
\includegraphics[angle=-90,scale=0.9]{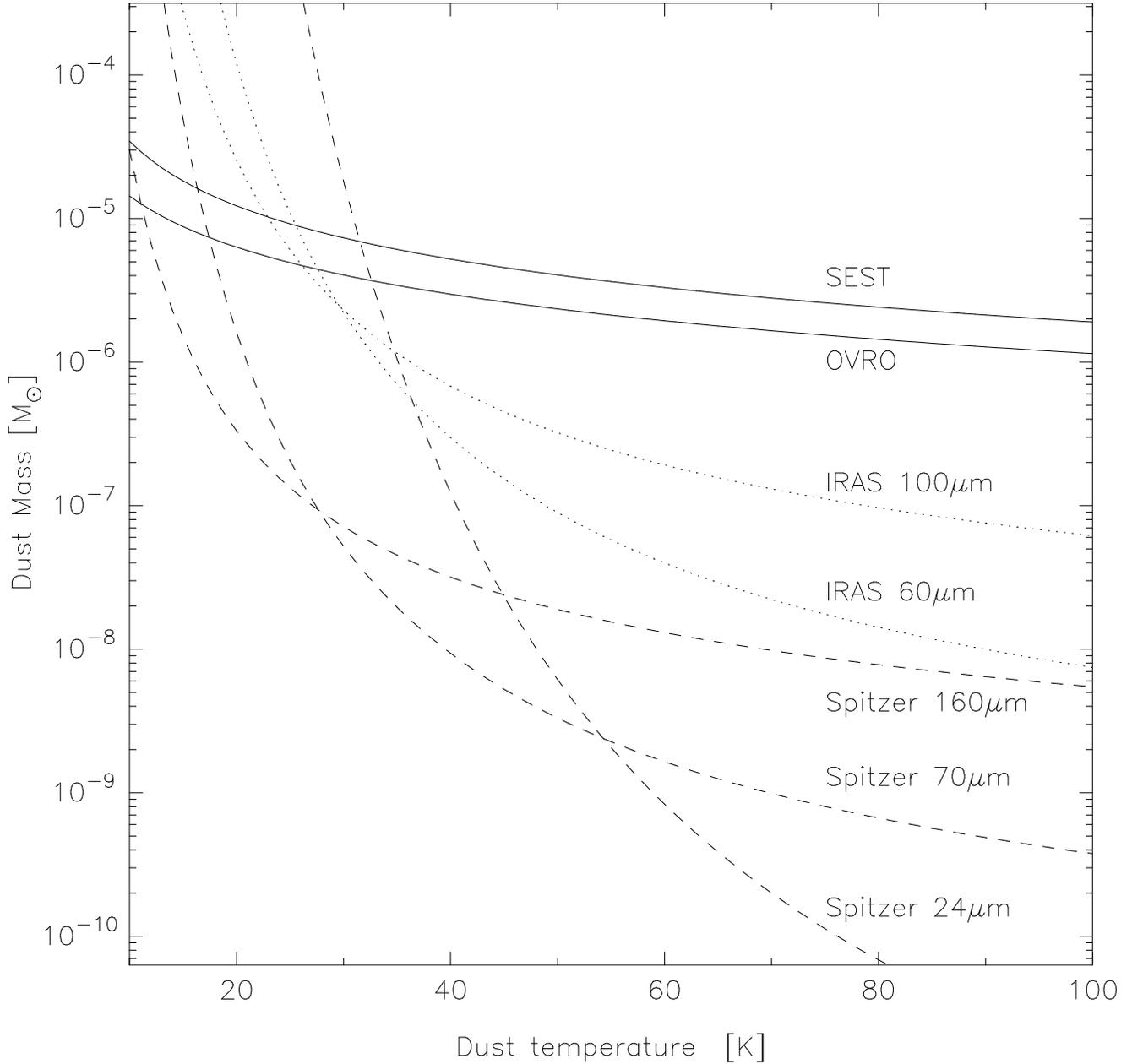}
\caption{
  \label{fig:tdust}
  The typical 5$\sigma$ sensitivity limits to dust mass as a function of the 
  dust temperature for the OVRO and SEST data (solid curves) compared 
  to {\it IRAS} (dotted curves) and {\it Spitzer} (dashed curves). Dust masses 
  were 
  computed for a source at a distance of 50~pc assuming that the emission is 
  isothermal and optically thin, with $\beta=1$ and 
  $\kappa(1.3{\rm mm})=0.02$\cmg. The millimeter continuum observations 
  presented in this paper are more sensitive to dust mass than {\it IRAS} for 
  temperatures colder than \about 25~K, and more sensitive than 
  {\it Spitzer} 24, 70, 
  and 160\micron\ for temperatures less than \about 30, 15, and 10~K 
  respectively. The {\it ISO} far-infrared photometric surveys 
  \citep{Habing01,Spangler01}, not shown here for clarity, have sensitities 
  intermediate between {\it IRAS} and {\it Spitzer}.
}
\end{figure}

In recent years, an increasing number of intermediate-aged, solar-type 
stars have been identified in nearby stellar associations and moving groups 
(see, e.g., Mamajek, Lawson, \& Feigelson~2000; Neuh\"auser \etal~2000; 
Mamajek, Meyer, \& Liebert~2002; Wichmann, Schmitt, \& Hubrig~2003; Song, 
Zuckerman, \& Bessell~2003). Some of these stars are being observed by the 
{\it Spitzer} Legacy program ``Formation and Evolution of Planetary Systems'' 
(or FEPS; Meyer \etal~2004) as part of a photometric and spectroscopic survey 
of \about 325 solar-type stars. The mid- and far-infrared {\it Spitzer}
observations will sensitive to relatively warm dust, and to survey for 
potentially cold dust around these stars, we conducted a 
millimeter-wavelength continuum survey of 125 stars in the FEPS sample.
Figure~\ref{fig:tdust} shows the sensitivity to dust mass of our millimeter 
continuum survey relative to that of {\it IRAS} and the upcoming FEPS 
{\it Spitzer} survey. As shown in this Figure, our observations are
more sensitive to dust mass than {\it IRAS} 60 and 100\micron\ data for dust
colder than \about 25~K, and more sensitive than {\it Spitzer} 24\micron, 
70\micron, 
and 160\micron\ observations for dust colder than \about 30~K, 15~K, and 10~K 
respectively. (Note, however, that the sensitivity limits for {\it IRAS} 
100\micron\ and {\it Spitzer} 70/160 160\micron\ observations depend on the 
local background level from cirrus.)
Accordingly, we have initiated a millimeter-wavelength continuum 
survey of a substantial fraction of the FEPS sample. The primary advantage of 
these observations over previous studies is that they sample a large number of 
stars (125) spanning a small range of stellar masses (0.5-2\msun) to 
specifically address the time scale for the evolution of cold circumstellar 
dust around solar-type stars.

In \S\ref{sample}, we summarize the sample selection and properties of
the FEPS stars observed for this study. The telescopes and instruments
used for these observations are described in \S\ref{obs}, along with the
data reduction procedures and the measured continuum fluxes. 
\S\ref{detections} summarizes the properties of the sources with detected 
millimeter continuum emission, and an analysis of the entire survey is
presented in \S\ref{analysis}. The implications for disk evolution are 
discussed in \S\ref{discussion}, and \S\ref{summary} summarizes our 
conclusions.

\section{Stellar Sample}
\label{sample}

The sources observed for this study were taken from the target list for the
FEPS {\it Spitzer} Legacy program \citep{Meyer04}, which consists of \about 
325 stars with stellar masses between \about 0.5\msun\ and 2\msun\ and ages 
spanning from \about 3~Myr to 3~Gyr. The FEPS source list includes field 
stars, stellar clusters (Pleiades, Alpha~Per, Hyades, and IC~2602), and
members of the Scorpius-Centaurus OB association. Ages were assigned based on 
a number of considerations, including: 
i) pre-main-sequence evolutionary tracks for stars with ages \aboutless 10~Myr,
ii) the observed lithium equivalent width, 
iii) x-ray activity, 
iv) the strength of the \ion{Ca}{2} H\&K emission for a volume limited sample
    of solar-type stars (\aboutless 50~pc), and 
v) the association of stars with clusters or star-forming regions of known age.
Distances were derived based upon Hipparcos parallaxes for nearby stars, 
kinematic distances for stars associated with young moving groups and 
associations (Mamajek \etal~2004, in preparation), and adopting nominal 
cluster distances for older clusters. Priority was placed on observing the 
youngest (\aboutless 300 Myr) and closest (\aboutless 100~pc) sources within 
the FEPS program. 

Table~\ref{tbl:sources} lists the 138 sources observed for this study along
with the adopted distances. We emphasize that the adopted distances and 
especially the assigned ages are preliminary, and a detailed analysis of the 
properties of the FEPS sample is forthcoming (Hillenbrand \etal~2004, in 
preparation). While all of these sources were initially in the FEPS target 
list, 13 sources were subsequently dropped due to time limitations or because 
ancillary ground-based observations cast doubt on the preliminary assigned age.
These 13 stars are listed separately in Table~\ref{tbl:sources} for
completeness and are otherwise not analyzed in this paper. None of the dropped 
sources were detected in the millimeter continuum with a signal to noise ratio
$\ge 3$. The 125 observed sources that are still 
included in the FEPS program are grouped into age bins that span a factor
of 3 as indicated in Table~\ref{tbl:sources}. The sources observed here 
include 14 stars between the ages of 3 and 10~Myr, 11 with ages of 
10-30~Myr, 39 with ages of 30-100~Myr, 39 with ages of 100-300~Myr, 
8 with ages of 300-1000~Myr, and 16 with ages of 1-3~Gyr. Submillimeter 
continuum photometry has been previously presented for five FEPS sources 
observed here: \citet{Williams04} detected 450\micron\ and 850\micron\ 
emission toward HD~107146, and and \citet{Liu04} report upper limits to the 
850\micron\ flux toward HD~35850, HD~199143, HD~129333, and HD~77407. 
In addition, 13 sources, as indicated in Table~\ref{tbl:sources}, have a 
$\ge3\sigma$ excess above the photospheric flux in one or more of the $IRAS$ 
bands as determined by fitting a Kurucz model atmosphere to compiled optical
and near-infrared photometry.

\section{Observations and Data Reduction}
\label{obs}

\subsection{OVRO}

Millimeter continuum observations of 57 stars were obtained in the summer 
of 2002 and the spring and summer of 2003 using the Owens Valley Radio 
Observatory (OVRO) millimeter-wave interferometer. The array contains six 10.4 
meter antennas, but during the summer months, only 3-5 antennas were typically 
operational at any one time. Data were obtained in either the OVRO ``compact'' 
or ``low'' configurations that provided a FWHM angular resolution of 5-10$''$. 
Continuum data were recorded simultaneously in four, 1~GHz wide continuum 
channels, and starting in the summer of 2003, low spectral resolution 
(32.5~MHz) data were also recorded using the COBRA spectrometer. At the time
of the observations, 7 of the 8 COBRA modules were functional, providing
7~GHz out of a possible maximum 8~GHz of bandwidth. When available, the COBRA 
spectral channels were averaged in 500~MHz intervals and used in place of the 
1~GHz continuum data during the data analysis. The majority of the 
observations were obtained at a mean frequency 
of 101~GHz as a compromise between achieving low system temperatures and 
atmospheric phase stability (favoring low frequencies) and dust emissivity 
(favoring high frequencies). Seven sources were observed at a mean frequency 
of 112~GHz as part of a survey for $^{12}$CO(J=1--0) molecular line emission; 
only the continuum data are presented here. A phase and 
amplitude calibrator was observed every 15 minutes, and the data were flux 
calibrated by observing Neptune or Uranus. 

The OVRO data were reduced using MIR, an IDL-based data reduction package for 
millimeter-wave interferometry developed by N. Scoville and J. Carpenter. 
Observations were repeated as necessary to achieve a typical RMS 
noise level of 0.5-1.0\mjybeam. To assess the calibration uncertainties, the 
dispersion of the measured fluxes was computed for each calibrator 
observed on 5 or more days within a 1 week time period (to reduce the 
likelihood of intrinsic source variability). Out of 161 calibrator sequences 
meeting these criteria, the median dispersion is 5\%, and 96\% of the 
sequences have a dispersion of less than 10\%. We adopt 5\% as the ``typical'' 
1$\sigma$ calibration uncertainty. Source fluxes were computed through least 
squares fitting of the visibility data assuming that the emission originates 
from a point source at the phase center. The reduced chi-square from the
fit was typically \about 1.8, where nominally it should be \about 1 if the
assigned uncertainties to the data points are correct and the point source
model is an adequate representation of the continuum emission. We assumed that 
the large value of the reduced chi-square indicates that the uncertainties 
in the visibility data have been underestimated, and accordingly we increased 
the uncertainties in the derived fluxes by the square root of the reduced 
chi-square value. Images were produced for each source using MIRIAD and 
visually inspected to search for continuum detections that may be offset from 
the phase center. None of the sources in this initial survey were detected at 
a signal to noise ratio $\ge$ 3.

During the course of this program, \citet{Williams04} detected 450\micron\ and 
850\micron\ continuum emission from HD~107146. This source was a non-detection 
in our original survey, but the submillimeter observations motivated us to 
obtain more sensitive, higher resolution 101~GHz data. These additional 
observations were obtained between 2003~April and 2004~May. We used 3C273 as 
the phase and gain calibrator, which is located at an angular separation of 
14.7$^\circ$ from HD~107146. Beginning in 2003~August, the radio source 
J1215+169, located 1.0$^\circ$ from HD~107146, was monitored after 
every 2 integrations on HD~107146 as a test of the accuracy of the phase 
calibration derived from 3C273. For most of these observations we were not 
able to observe a planet to directly flux calibrate the data, and instead we
used 3C273 as a secondary flux calibrator. Since the millimeter flux from 
3C273 is frequency dependent and time variable (see, e.g., Steppe \etal~1993),
we determined the flux history of 3C273 using observations obtained by other
programs at OVRO. Using 201 OVRO ``tracks'' in which both 3C273 and a planet 
(either Neptune or Uranus) were observed in the time frame of the HD~107146 
observations, we found that the flux from 3C273 varied in time between 
7 and 10~Jy at observed frequencies of 86 to 115~GHz. We determined that the 
flux from 3C273 varied as $S_\nu \propto \nu^{-0.9}$ by fitting a power law 
to a portion of the data where the 3C273 flux was \about 10~Jy and appeared to 
have low variability. While the spectral index can vary between \about 0.3
and 1.5 during 3C273 flare events, our derived spectral index as well as the 
absolute flux level suggests 3C273 was largely in a quiescent phase for the 
time period of our observations \citep{Robson93}. Therefore, we assumed that 
the spectral index was constant. After scaling the 
observed fluxes to 101~GHz, a median boxcar curve, with a width corresponding
to 15 observed data points, was fitted through the resulting time series data 
to estimate the flux for 3C273 for any given time. The residuals from the 
fitted curve have a RMS of 6\%, which we adopt as the calibration 
uncertainty in the measured fluxes for 3C273 and consequently HD~107146.
The HD107146 data were reduced in MIR, and images were produced and cleaned in 
MIRIAD using a Brigg's robust weighting parameter \citep{Briggs95} of 0. The 
resulting image had a FWHM synthesized beam of $4.5''\times4.0''$ and a RMS 
noise of 0.15~\mjybeam. Analysis of the HD~107146 data is presented in 
\S\ref{hd107146}.

\subsection{SEST}

Continuum observations at $\lambda$1.2~mm were obtained for 89 stars using the 
37-element SIMBA bolometer camera on the 15~m Swedish-ESO Submillimetre 
Telescope (SEST) in 2001 November, 2002 June, and 2002 November. The 
instantaneous field of view of SIMBA is \about $4'\times4'$ with 44\arcsec\ 
separation between the bolometer channels. The FWHM beam size of the 
observations is 24\arcsec. Fully sampled maps were generated by dithering the 
array to produce images with 8$''$ sampling. The telescope pointing was 
checked before and after each map, and the pointing offsets were repeatable 
to within 2$''$ in both azimuth and elevation in most cases. However, 
for the observations of both RX~J1852.3$-$3700 and MML~1 in June~2002, the 
pointing offsets derived for the start and end of the maps differed by 
\about 40$''$ in elevation. A continuum source was detected in the 
RX~J1852.3$-$3700 map (see \S\ref{photometry} and \S\ref{rxj1852}) that was 
offset 40$''$ from the pointing center. We assumed that the 40$''$ angular 
offset was due to the pointing error and associate this continuum source with 
RX~J1852.3$-$3700. The atmospheric opacity at $\lambda$1.2~mm as 
measured from sky dips ranged from 0.12 to 0.20 in 2002~June and 0.25 to 0.35 
in 2001~November and 2002~November. Flux calibration was derived from maps of
Uranus and Neptune which were obtained each night. The estimate 1$\sigma$
calibration uncertainty is 10\% based on repeated observations of the planets.
Source fluxes were measured using a beam-weighted aperture \citep{Naylor98} of 
16$''$ in radius with a background annulus that extended from 30$''$ to 
120$''$. If no source was apparent in the images, the aperture was centered on 
the stellar position. The typical RMS noise measured in the background annulus 
varied from \about 8\mjybeam\ for the 2002 June observations to 17\mjybeam\ 
for the 2001~November and 2002~November observations.

\subsection{CSO}

Submillimeter continuum observations at $\lambda350$\micron\ were obtained 
for 6 stars using the SHARC bolometer camera \citep{Hunter96} on the 10.4~m 
telescope of the Caltech Submillimeter Observatory (CSO). The observations
were conducted during the second half of the night on 2001 October 1-3. SHARC 
contains a 24 pixel monolithic silicon bolometer array, of which 19 pixels were 
operational at the time of the observations. Each pixel subtends 
$5''\times10''$ on the sky. The FWHM beam size of CSO telescope at the
$\lambda$350\micron\ is \about 9$''$. Data were obtained in 
point-source mode by chopping 1\arcmin\ in azimuth, with on-source integration 
times ranging from 12 to 36 minutes. Zenith opacities at 225~GHz as measured 
by the CSO tau meter ranged between 0.04 and 0.09. The observations were flux 
calibrated with regular observations of Saturn, with an estimated 1$\sigma$
calibration uncertainty of 20\% based on the repeated observations. Pointing 
was checked every 1-2 hours, where it was discovered that there was an 
elevation pointing drift as a function of elevation for the first two nights 
before the pointing model was updated on the last night. During reduction, the 
data were combined using a shift-and-add procedure to account for the drift. 
The RMS noise in the calibrated, coadded data ranged from 37 to 57\mjybeam. 

\subsection{Photometry}
\label{photometry}

\begin{figure}
\epsscale{0.70}
\includegraphics[angle=-90,scale=0.7]{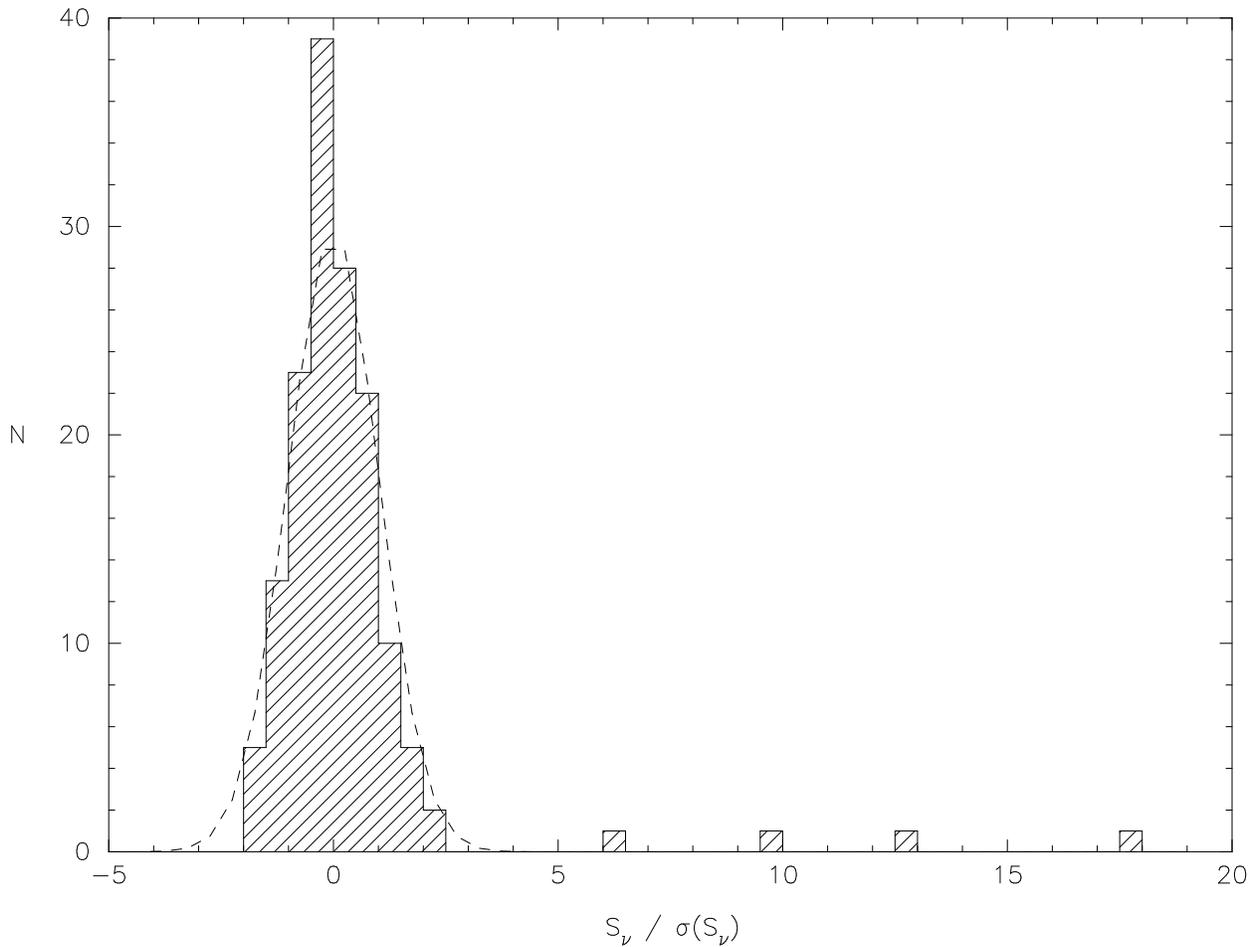}
\caption{
  \label{fig:snr}
  Histogram of the observed submillimeter and millimeter continuum fluxes 
  normalized by the internal measurement uncertainties. The dashed curve 
  shows the expected distribution of this ratio for the number of observed
  stars assuming random gaussian noise. Four sources (RX~J1842.9$-$3532, 
  RX~J1852.3$-$3700, PDS~66, and HD~107146) have been detected with a signal
  to noise ratio greater than $3\sigma$, while the remaining sources
  have fluxes consistent with random noise.
}
\end{figure}

The measured millimeter and submillimeter fluxes for all observed sources are
summarized in Table~\ref{tbl:sources}. With the exception of HD~107146, 
which was resolved (see \S\ref{hd107146}), the fluxes are for point sources
at the stellar position. These flux measurements then do not account for
any extended continuum emission. For cases where large pointing 
uncertainties were present and the stellar position in the image is uncertain 
by more than one beam width, the table lists the flux as the 3$\sigma$ upper 
limit, which was computed as 3 times the RMS noise in the image. The flux
uncertainties listed in Table~\ref{tbl:sources} represent the internal
uncertainties only and do not include the calibration uncertainties.
Figure~\ref{fig:snr} presents a histogram of the ratio of observed flux to the
internal flux uncertainty. The dashed curve shows the expected frequency 
distribution of this ratio for gaussian noise; i.e. a normal distribution 
centered on zero with unit dispersion. Evidently gaussian noise is an adequate 
representation of the observed flux distribution for most sources. Four 
sources were detected with a signal to noise ratio $\ge 3$:
HD~107146 at $\lambda$3.0~mm with the OVRO interferometer, and PDS 66, 
RX~J1842.9$-$3532, and RX~J1852.3$-$3700 at $\lambda$1.2~mm with the
SEST bolometer. As discussed in \S\ref{hd107146}, the emission toward 
HD~107146 is resolved
with OVRO. For the 3 sources detected with SEST, the observed FWHM of the 
continuum emission, measured by fitting a gaussian to the emission radial 
profile using the IMEXAMINE task in IRAF, varies between 24-28$''$.
Given that the FWHM beam size of the SEST telescope is 24$''$, the observed 
emission is consistent with a point source at this angular resolution.

\section{Detected Millimeter Continuum Sources}
\label{detections}

\begin{figure}
\epsscale{0.90}
\includegraphics[angle=-90,scale=0.7]{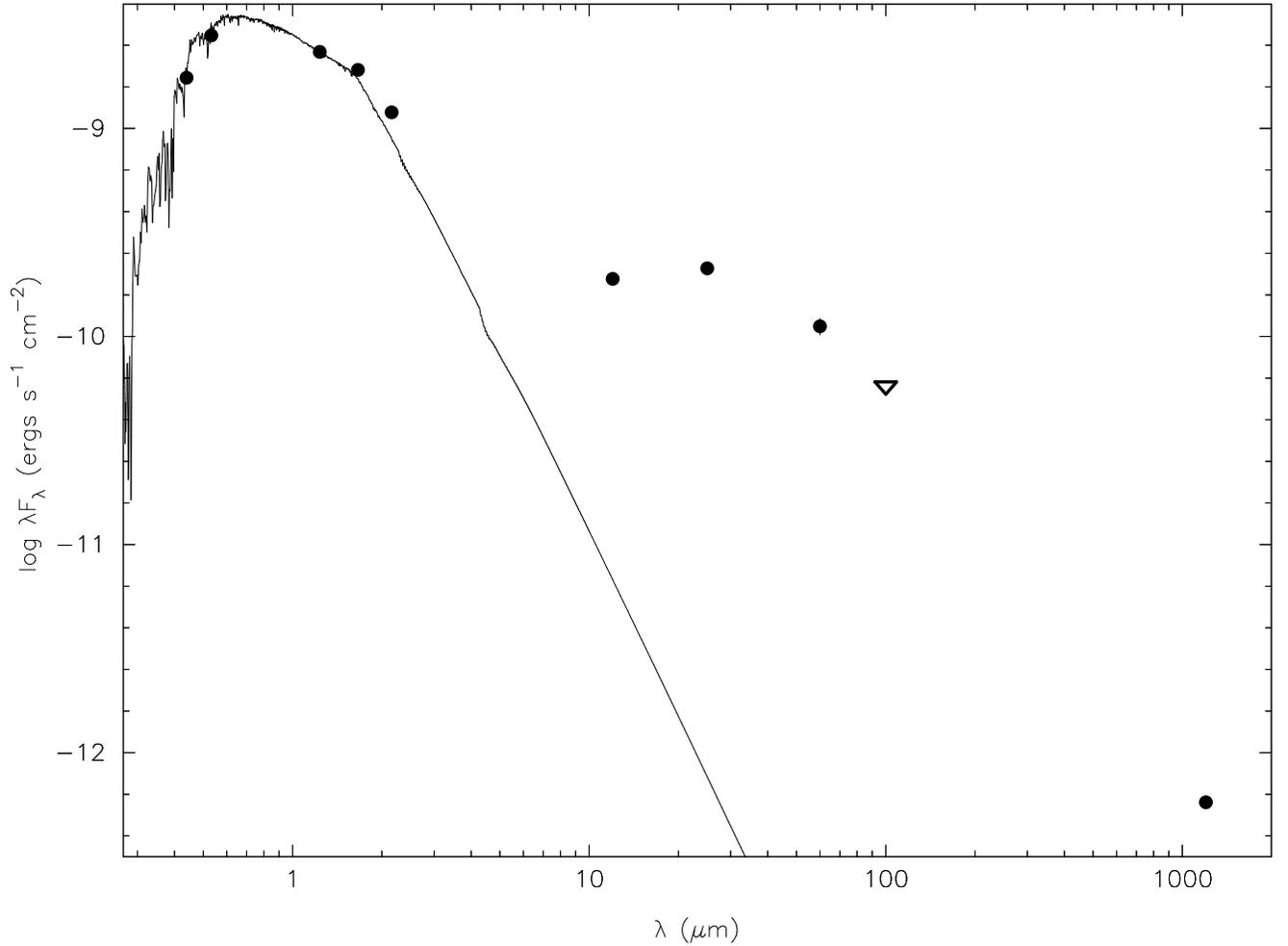}
\caption{
  \label{fig:pds66}
  The spectral energy distribution for PDS 66, a \about7-17~Myr old star in 
  the Lower Centaurus-Crux subgroup of the Scorpius-Centaurus OB association
  \citep{Mamajek04}. Filled circles indicate detected broad-band fluxes
  from Tycho $(BV)_T$ \citep{Hog00}, 2MASS $JHK_s$, and {\it IRAS}, and the 
  open triangle represents an upper limit from {\it IRAS}. The solid line is 
  the best fit Kurucz model of $T_{eff}=5150$~K and $A_{V}=0.80^{\rm m}$.
}
\end{figure}

\begin{figure}
\epsscale{0.90}
\includegraphics[angle=-90,scale=0.7]{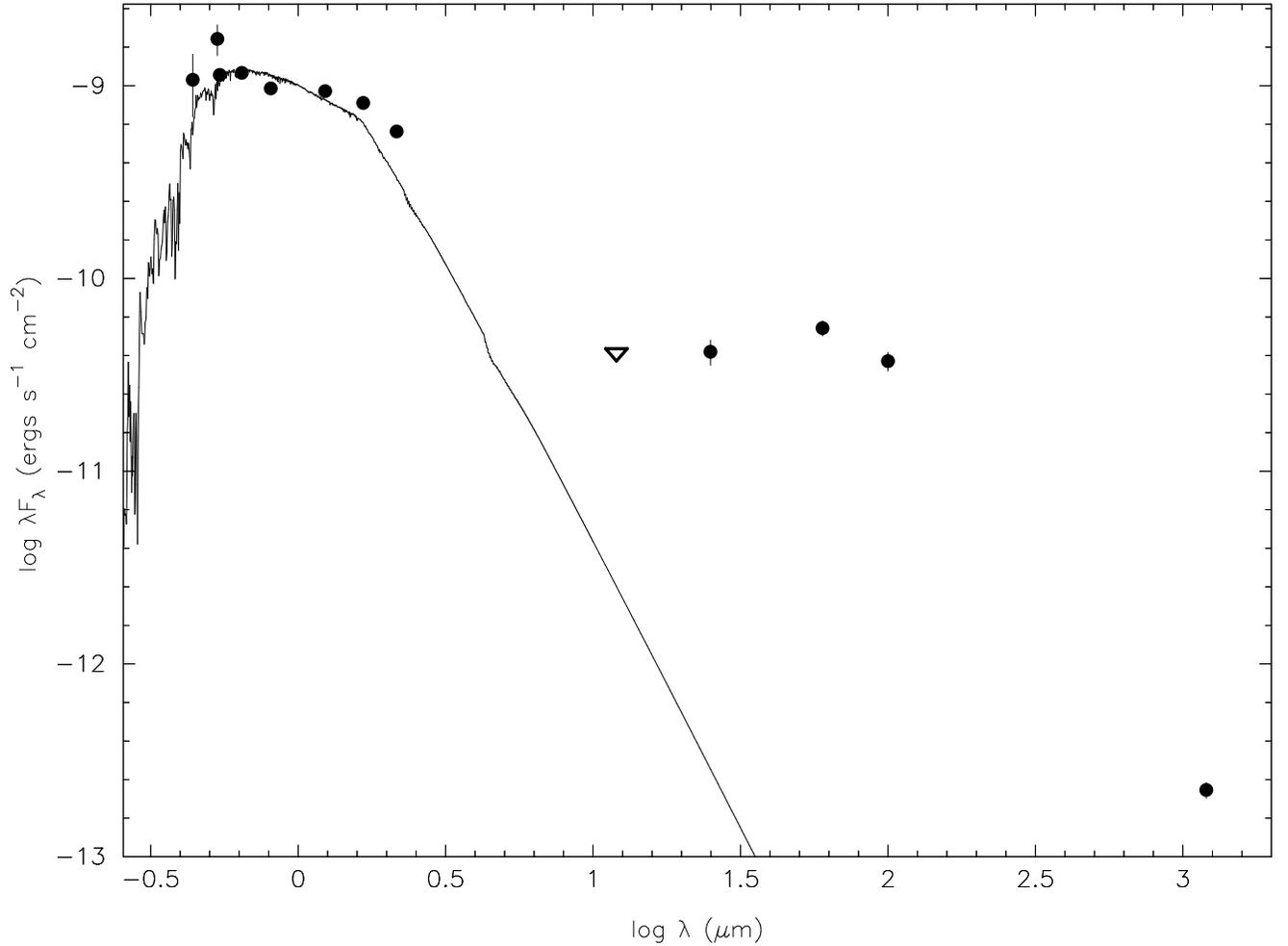}
\caption{
  \label{fig:rx1842}
  The spectral energy distribution for RX J1842.9$-$3532, a \about 10~Myr
  star that is located in projection nearby the RCrA molecular cloud
  \citep{Neuhauser00}. Filled circles indicate detected 
  broad-band fluxes from Tycho $(BV)_T$
  \citep{Hog00}, $V(RI)_C$ \citep{Neuhauser00}, 2MASS $JHK_s$, and {\it IRAS},
  and the open triangle is an upper limit from {\it IRAS}.
  The solid curve is the best fit Kurucz model of $T_{eff}=5050$~K and 
  $A_{V}=1.6^{\rm m}$.
}
\end{figure}

\begin{figure}
\epsscale{0.90}
\includegraphics[angle=-90,scale=0.7]{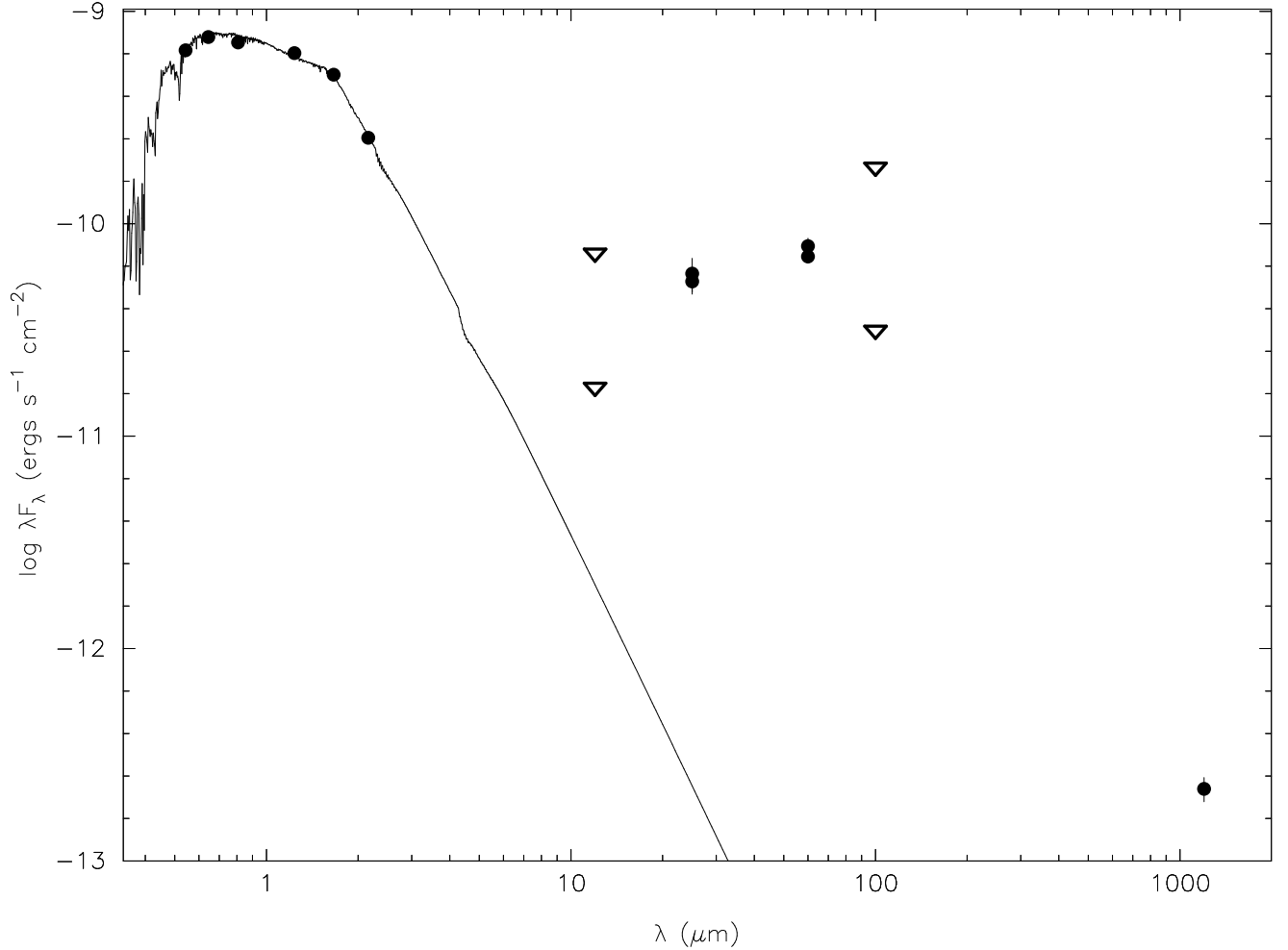}
\caption{
  \label{fig:rx1852}
  The spectral energy distribution for RX J1852.3$-$3700, a \about 10~Myr
  star that is located in projection nearby the RCrA molecular cloud
  \citep{Neuhauser00}. Filled circles indicate detected
  broad-band fluxes from Tycho $(BV)_T$ \citep{Hog00}, $V(RI)_C$ 
  \citep{Neuhauser00}, 2MASS $JHK_s$, and {\it IRAS}, and the open triangles
  indicate upper limits from the IRAS Point Source and Faint Source catalogs.
  The solid curve is the best fit Kurucz model of $T_{eff}=4800$~K and 
  $A_{V}=1.1^{\rm m}$.
}
\end{figure}

Figures~\ref{fig:pds66}--\ref{fig:rx1852} show the observed spectral energy 
distribution (SED) for the 3 stars detected with SEST. The SED for HD~107146 
is presented and analyzed by \citet{Williams04} and is not shown here.
In addition to the millimeter-wavelength continuum photometry, the SEDs
include photometry from Tycho-2 \citep{Hog00}, {\it 2MASS}, {\it IRAS}, 
and $V(RI)_C$ for two sources \citep{Neuhauser00}. Also shown in these 
figures is the best fit Kurucz model atmosphere to the optical and 
near-infrared photometry. The best-fit model was obtained by minimizing the 
chi-square between the model magnitudes and observed $(BV)_TV(RI)_C$ and $JH$ 
photometry, where the free parameters in the fit are the overall normalization 
constant, the effective temperature, and visual extinction. Model magnitudes 
were computed by convolving the Kurucz model with the appropriate filter 
transmission as described in Cohen \etal~(2003a,b) and references therein. 
Each of these sources exhibits excess emission in the {\it IRAS} bands,
and both PDS~66 and RX~J1842.9$-$3532 may have excess emission at wavelengths
as short as 2.2\micron. An analysis of the SEDs is not presented
here since the FEPS project will soon have {\it Spitzer} spectrophotometry 
from 3.6 to 160\micron\ for these sources that will permit significantly more 
detailed circumstellar disk models to be constructed than is possible with the 
limited data available now. The following discussion briefly summarizes the 
properties of the four sources detected in the millimeter continuum.

\subsection{PDS 66}
\label{pds66}

PDS~66 (also known as Tycho~9246~971~1, Hen~3-892, IRAS~13185-6922, and MML~34)
was first identified by \citet{Gregorio92} as a classical T Tauri star (CTTS)
by virtue of having a far-infrared excess detected by {\it IRAS}, strong 
H$\alpha$ emission, and deep lithium absorption at $\lambda$6707\AA.
\citet{Mamajek02} identified the star as a likely proper motion member of the 
Lower Centaurus-Crux subgroup of the Scorpius-Centaurus OB association
and derived the stellar parameters, which we adopt here. The
star has a spectral type of K1~IVe, and is distinguished as the only known 
CTTS in a sample of 110 stars in the Lower Centaurus-Crux and 
Upper Centaurus-Lupus subgroups. Depending on the pre-main-sequence 
evolutionary tracks used to estimate the stellar properties, the stellar age 
and mass are 7-17~Myr and 1.1-1.2\msun, respectively. The mean age of the 
Lower Centaurus-Crux subgroup is 17-23~Myr. We adopt the secular parallax 
distance of 86 pc as derived by \citet{Mamajek02}. The observed 
millimeter continuum flux corresponds to a dust mass (see \S\ref{discussion}) 
of \about $5\times10^{-5}$\msun, which is comparable to the circumstellar disk 
masses found around young stars in the Taurus \citep{Beckwith90,Osterloh95} 
and $\rho$~Oph \citep{Andre94,Motte98} molecular clouds. Thus PDS~66 appears 
to be a rare example of a relatively old (\aboutmore 10~Myr) CTTS, and the 
millimeter continuum emission likely originates from a circumstellar accretion 
disk.

\subsection{RX J1842.9$-$3532}
\label{rxj1842}

As described in \citet{Neuhauser00}, the x-ray source RX~J1842.0$-$3532 was 
detected in the {\it ROSAT} All-sky survey and subsequently associated with 
the star Hen~3-1722 and IRAS 18396$-$3535. The star has a K2 spectral type, 
exhibits strong H$\alpha$ in emission (see also Henize~1976, Beers \etal~1996), 
has a deep lithium absorption feature, and contains excess {\it IRAS} 
far-infrared emission. Based on these characteristics the star has been 
classified as a CTTS. RX~J1842.0$-$3532 is located at an angular separation of
\about 3.6$^\circ$ from the RCrA molecular cloud. The observed radial 
velocities and UCAC2 proper motions for the star agree to within 1~km~s$^{-1}$ 
of the values predicted for RCrA members at their positions
(using the space motion vector for the RCrA group from Mamajek \& 
Feigelson~2001; E.Mamajek, private communication). The secular parallax 
distance is statistically consistent with the best available distance 
of 130~pc to the RCrA association \citep{Casey98}. For this distance, 
\citet{Neuhauser00} estimate a stellar age of \about 10~Myr and a stellar mass 
of 1.2\msun\ based on pre-main-sequence evolutionary tracks. The observed
millimeter continuum flux implies a dust mass (see \S\ref{discussion}) of 
\about $5\times10^{-5}$\msun. Given that the star has characteristics similar 
to that of a CTTS, the dust emission likely originates from an accretion disk.

\subsection{RX J1852.3$-$3700}
\label{rxj1852}

As described in \citet{Neuhauser00}, the x-ray source RX~J1852.3$-$3700 was 
detected in the {\it ROSAT} All-sky survey and associated with a stellar 
counterpart that has strong H$\alpha$ emission and far-infrared emission as 
detected by {\it IRAS} (IRAS 18489$-$3703). The star has a K7 spectral type and 
has been classified as a CTTS. It is located at an angular separation of 
\about 1$^\circ$ from the RCrA molecular cloud, and similar to 
RX~J1842.9$-$3532, the observed radial velocities and UCAC2 proper motions
agree to within 1~km~s$^{-1}$ to the predicted values for RCrA members
(E. Mamajek, private communication). For the adopted distance of 130~pc to
RCrA \citep{Casey98}, the estimated stellar age and mass from 
pre-main-sequence evolutionary tracks are $\sim$10~Myr and 1.1\msun\
respectively \citep{Neuhauser00}. The observed millimeter continuum flux 
implies a dust mass (see \S\ref{discussion}) of \about $5\times10^{-5}$\msun. 
Since the star has characteristics similar to that of a CTTS, the dust 
emission likely originates from a circumstellar accretion disk.

\subsection{HD 107146}
\label{hd107146}

HD~107146 has a G2~V spectral type with an Hipparcos distance estimate of 
29~pc. The star has a number of age indicators (lithium equivalent width, 
x-ray luminosity, space motions similar to Pleiades moving group) which in 
aggregate suggest an age of \about 80-200~Myr (see discussion in 
Williams \etal~2004). \citet{Silverstone00} and \citet{Metchev04} noted that 
the source contains an {\it IRAS} far infrared excess at 60\micron\ and 
100\micron. \citet{Williams04} further detected continuum emission at 
450\micron\ and 850\micron\ that firmly established that the source is 
surrounded by circumstellar dust, most likely originating from a 
debris disk. Their analysis of the SED suggested that the disk contains an
inner hole with a radius $>$ 31~AU.

\begin{figure}
\epsscale{0.90}
\includegraphics[angle=-90,scale=0.7]{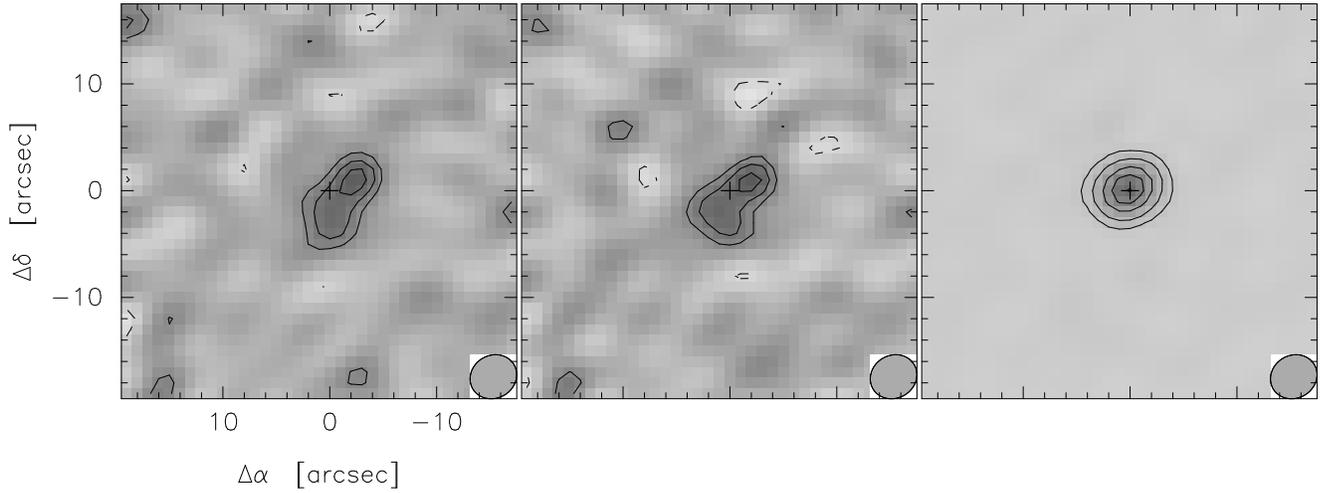}
\caption{
  \label{fig:hd107146}
  {\it Left panel:} OVRO $\lambda$3~mm continuum cleaned images of HD~107146 
  produced using all available data with robust=0 weighting \citep{Briggs95}. 
  The RMS noise in the image 
  is 0.15\mjybeam\ with a synthesized FWHM beam size of $4.5''\times4.0''$ as 
  shown in the lower right corner of each panel. The solid 
  contours begin at $2\sigma$ with increments of $1\sigma$, and the dashed 
  contours start at $-2\sigma$ with increments of $-1\sigma$. The plus (+)
  symbol at (0,0) indicates the stellar position.
  {\it Middle panel:} Same as the left panel, except for the subset of the 
  data where J1215+169 was monitored to assess the accuracy of the phase 
  calibration. The RMS noise in the map is 0.16\mjybeam.
  {\it Right panel:} Contour map of J1215+169, which was observed
  along with the data shown in the middle panel to empirically test the
  image quality. Contours are drawn at 10, 25, 50, and 75\% of the peak
  flux (0.37~Jy).
}
\end{figure}

Figure~\ref{fig:hd107146} presents the OVRO $\lambda$3~mm map of HD~107146. 
The left panel shows the contour map of the continuum emission
obtained by combining all available data. The middle panel displays the 
continuum image for a subset of the data when J1215+169 was monitored 
as a test of the phase/gain solution derived from 3C273, and the right panel 
shows the corresponding image of J1215+169. The observed peak flux in the 
HD~107146 image is $0.69\pm0.15$\mjybeam\ in a $4.5''\times4.0''$~beam. The 
integrated flux in a 8$''$ radius aperture, measured in a naturally-weighted 
map ($6.7''\times6.0''$ beam size) to optimize the signal to noise, is 
$1.42\pm0.23$~mJy. (The uncertainties in the peak and integrated fluxes do not 
include the 6\% calibration uncertainty; see \S\ref{obs}.) \citet{Williams04} 
fitted the excess infrared and submillimeter emission with a graybody to
derive a dust temperature of 51$\pm$4~K, and estimated the frequency 
dependence of the submillimeter emission to be 
$S_\nu \propto \nu^{2.69\pm0.15}$ in the Rayleigh-Jeans limit. Based on the 
observed 850\micron\ flux of $20\pm4$~mJy, the expected 3~mm flux for these 
model parameters is $0.78\pm0.21$~mJy. The difference between the observed 
and expected 3~mm flux, including calibration uncertainties, is 
$0.64\pm0.32$~mJy, or a 2.0$\sigma$ difference. Thus the observed OVRO 3~mm 
flux is marginally consistent with that expected from the graybody 
extrapolation of the submillimeter emission.

The centroid position of the continuum emission toward HD~107146, measured
by fitting an elliptical gaussian to the visibility data, is offset by
($-0.46''\pm0.45,-0.51''\pm0.43$) from the stellar position. (The uncertainty 
in the stellar position, including uncertainties in the Hipparcos astrometry
and Tycho-2 proper motions, are \about 0.013$''$ for both the
right ascension and declination.) The astrometric uncertainty due to 
phase-correction uncertainties is $<0.35''$ as determined from the measured 
offset of J1215+169 from the phase center (see Fig.~\ref{fig:hd107146}). Thus 
within the astrometric uncertainties, the centroid of the millimeter continuum 
emission is centered on the stellar position. \citet{Williams04} indicated 
that the peak 450\micron\ emission may be offset by 4.4$''$ from the stellar 
position. This offset is inconsistent with the OVRO observations assuming 
that the 450\micron\ and 3~mm emission arise from the same location in the 
disk.

\begin{figure}
\epsscale{0.80}
\includegraphics[angle=-90,scale=0.9]{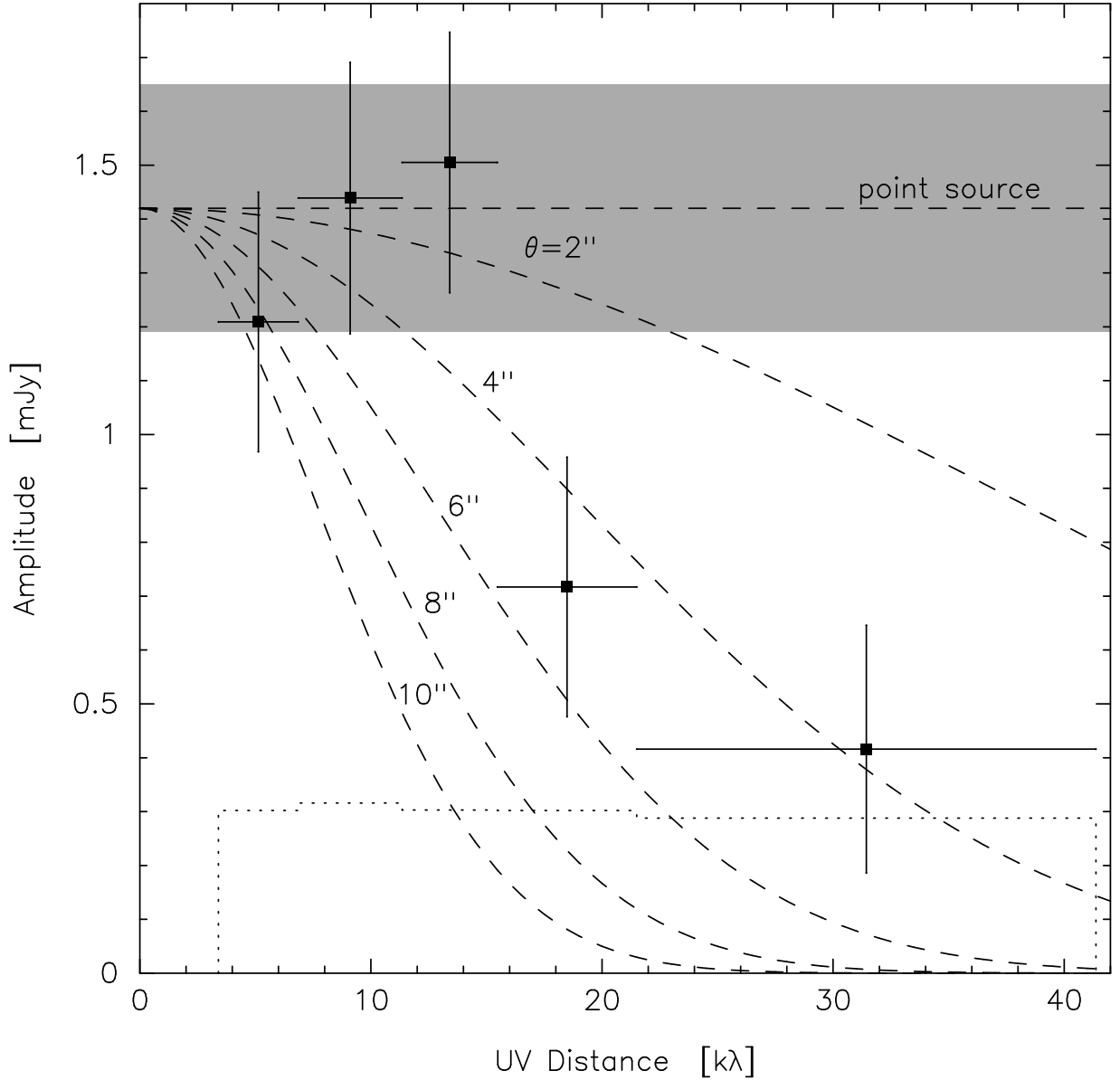}
\caption{
  \label{fig:amp}
  The vector-averaged visibility amplitude (solid squares) as a function of 
  $uv$ distance for the OVRO $\lambda$3~mm observations of HD~107146. The 
  uncertainties in the amplitudes represent the standard deviation of the 
  mean of the visibility data, while the horizontal lines through the data
  points reflect the full-width of the $uv$ bin. The data were binned to
  provide an equal number of $uv$ points per bin. The dotted line shows the 
  expectation value of the visibility amplitudes in the absence of any source.
  The gray shaded region shows the observed integrated flux of
  $1.42\pm0.23$~mJy as measured in the image domain. The dashed lines show the 
  expected visibility amplitudes as a function of $uv$-distance for noiseless
  point source and circular gaussian (with FWHM between $2''$ and 10$''$) 
  models that have an integrated intensity of 1.42~mJy. The decrease in the 
  visibility amplitudes for $uv$ distances greater than \about 15~k$\lambda$ 
  indicates that the source has been resolved at a significance level of 
  3.8$\sigma$. The best-fit elliptical gaussian to the visibility data has a 
  FWHM size of $(6.5''\pm1.4'') \times (4.2''\pm1.3'')$.
}
\end{figure}

Figure~\ref{fig:hd107146} suggests that the continuum emission around HD~107146
may be resolved with the OVRO interferometer in that the lowest contours are 
extended at a position angle (east of north) of approximately $-55^\circ$, 
which is not observed in J1215+164. To quantify the possible extension in the 
continuum emission, Figure~\ref{fig:amp} shows the average observed visibility 
amplitude for the HD~107146 data as a function of $uv$ distance from the phase 
center. The phase center was adopted as the reference point since within the 
astrometric uncertainties the continuum emission is centered on the stellar
position as discussed above. The visibility data were averaged in 5 $uv$ bins,
where the bin sizes vary as a function of $uv$ distance to maintain a constant
number of visibility points per bin. The amplitude uncertainties 
represent the standard deviation of the mean in the visibility data, and the 
horizontal lines through each data point represent the full width of the $uv$ 
bin. The dashed curves show the expected amplitudes as a function of $uv$ 
distance in the absence of noise for model circular gaussians that have 
integrated intensities of 1.42~mJy and FWHM ranging from $\theta=0''$ (i.e. a 
point source) to 10$''$. The shaded region indicates the $\pm1\sigma$ 
amplitude uncertainty of the integrated flux measured in the image domain.

As Figure~\ref{fig:hd107146} shows, the last two radial bins are below the 
expected amplitudes for a point source model. The vector averaged amplitude 
of the visibility data in these two bins combined is $0.33\pm0.17$~mJy, such 
that the deviation from the point source amplitude flux is $1.1\pm0.29$~mJy, 
or a 3.8$\sigma$ deviation. We thus conclude that the emission toward 
HD~107146 has indeed been resolved. To estimate the source size, an elliptical 
gaussian was fitted to the unbinned visibility data points. The position
angle of the best-fit gaussian is $-34^\circ\pm23^\circ$, which agrees well
with the direction of extended emission at 450\micron\ noted by 
\citet{Williams04}. The fitted FWHM gaussian size is 
$(6.5''\pm1.4'') \times (4.2''\pm1.3'')$, which corresponds to a spatial size 
of \about $185~{\rm AU}\times120~{\rm AU}$. Assuming that the disk is 
intrinsically circularly symmetric, the ratio of the minor to major axis 
implies an inclination angle of $50^\circ\pm18^\circ$ (where $0^\circ$ is
for a face on disk). The derived FWHM size is smaller than the 
$10.5''\times7.4''$ size inferred from the 450\micron\ emission, which is 
unexpected since both the 450\micron\ and 3mm emission are likely 
optically thin and in the Rayleigh-Jeans limit \citep{Williams04}. To have a 
smaller $\lambda$3~mm size, one needs a grain population in the disk interior 
that radiates more effectively at longer wavelengths. In principle this could 
signify the presence of larger grains which would be in radiative equilibrium 
at colder temperatures. While in fact the observed $\lambda$3mm flux is 
slightly larger than the extrapolated submillimeter flux, the moderate signal 
to noise ratio for both the $\lambda$450\micron\ and $\lambda$3mm images
are such that higher signal to noise detections are needed before drawing firm 
conclusions on any size differences as a function of wavelength.

Finally, we examine the consistency of the debris disk size with the 
characteristic dust temperature of $51\pm4$~K inferred from the SED analysis.
For $\beta$=0.69 (see \S\ref{dustmass}) and assuming a single grain size that 
efficiently absorbs the stellar radiation, the radial dependence of the dust 
temperature around a solar luminosity star will be 
$T_{\rm dust} \approx 412{\rm K}~r_{\rm AU}^{-0.43} L_*^{0.21} a_{\mu\rm m}^{-0.15}$,
where $r_{\rm AU}$ is the orbital radius measured in AU, $L_*$ is the 
luminosity in solar luminosities, and $a_{\mu\rm m}$ is the grain size in 
microns \citep{Backman93}. The resolved disk implies a range of dust 
temperatures must be present. At the half-width radius of 90~AU, the dust 
temperature will be 60~K for 1\micron\ sized grains with warmer dust
at small radii interior. These are warmer temperatures than implied by the
SED modeling, but can be reconciled by invoking larger grains a few microns 
in size that radiate more efficiently and result in cooler temperatures. In a 
future FEPS publication, the HD~107146 debris disk will be presented that 
incorporates the submillimeter photometry, disk size, and {\it Spitzer} data
to present a derive a self-consistent model.

\section{Analysis}
\label{analysis}

\subsection{Dust Masses}
\label{dustmass}

The observed millimeter continuum fluxes can be used to estimate, or place
limits on, the circumstellar dust masses. Assuming that the emission is 
isothermal and optically thin, the dust mass was computed using the following
formula
\begin{equation}
M_{dust} = {S_\nu\:D^2 \over \kappa_\nu\:B_\nu(T_{dust})},
\label{eq:mass}
\end{equation}
where $\kappa_\nu = \kappa_o({\nu\over\nu_o})^\beta$ is the mass absorption
coefficient, $\beta$ parameterizes the frequency dependence of $\kappa_\nu$,
$S_\nu$ is the observed flux, $D$ is the distance to the source, $T_{dust}$ 
is the dust temperature, and
$B_\nu(T_{dust})$ is the Planck function \citep{Hildebrand83}. We assumed 
$\beta$=1.0 and $\kappa_o$ = 2\cmg\ at $\lambda$1.3~mm \citep{Beckwith90}, and 
adopted a dust temperature of 40~K, which is a compromise between the expected 
cold (\about 20-30~K) dust in optically 
thick accretion disks \citep{Beckwith90,Andre94}, and the warmer dust 
(45-100~K) inferred for optically thin debris disks around solar type stars
\citep{Zuckerman04}. These mass estimates reflect the mass contained in small 
dust grains, and do not account for mass contained in large (millimeter-sized) 
particles that do not contribute significantly to the emission at millimeter 
wavelengths. For these adopted parameters, the inferred dust masses for the 
detected sources are $3.2\times10^{-7}$\msun, $5.0\times10^{-5}$\msun, 
$5.1\times10^{-5}$\msun, and $5.0\times10^{-5}$\msun\ for HD~107146, PDS~66, 
RX~J1842.9$-$3532, and RX~J1852.3$-$3700 respectively.

A number of sources of systematic errors are present in computing in the dust 
masses, a few of which we mention here. Since the dust mass is proportional 
to $T_{\rm dust}^{-1}$ in the Rayleigh-Jeans limit, the dust masses will be 
uncertain by a factor of \about 2 due to the dust temperature alone. Further,
the value for $\beta$ can vary among sources over the range of \about 0 to
1.5, with most values \aboutless 1 \citep{Beckwith91,Weintraub89} most common.
Adopting a value of $\beta=0.5$ for example would decrease the masses derived
from the OVRO $\lambda$3mm observations by a factor of 1.5. A larger source of 
uncertainty in computing the dust masses is the value for the dust opacity, 
$\kappa_\nu$ (see review by Beckwith, Henning, \& Nakagawa~2000). For example, 
\citet{Pollack94} computed $\kappa_\nu$ for a variety of grain sizes and 
compositions, and found values that ranged from 0.14 to 0.87\cmg\ at 
$\lambda$1.3~mm for 0.1\micron-3~mm radius grains (see also Stognienko, 
Henning, \& Ossenkopf~1995). The commonly used value for debris disks of 
1.7\cmg\ at $\lambda800$\micron, adopted by \citet{Zuckerman93} to place a 
lower limit on the amount of dust, corresponds to 1.0\cmg\ at $\lambda1.3$~mm 
assuming $\beta=1$. Given that $\kappa_o$ is poorly constrained, the relative 
values of disk masses should be significantly more reliable than the absolute 
values. We recognize, however, that systematic changes in $\kappa_\nu$ with 
age are likely present, and therefore any trends of dust mass with stellar age 
can be interpreted as variations in dust mass and/or grain properties. 

\begin{figure}
\epsscale{0.80}
\includegraphics[angle=-90,scale=0.7]{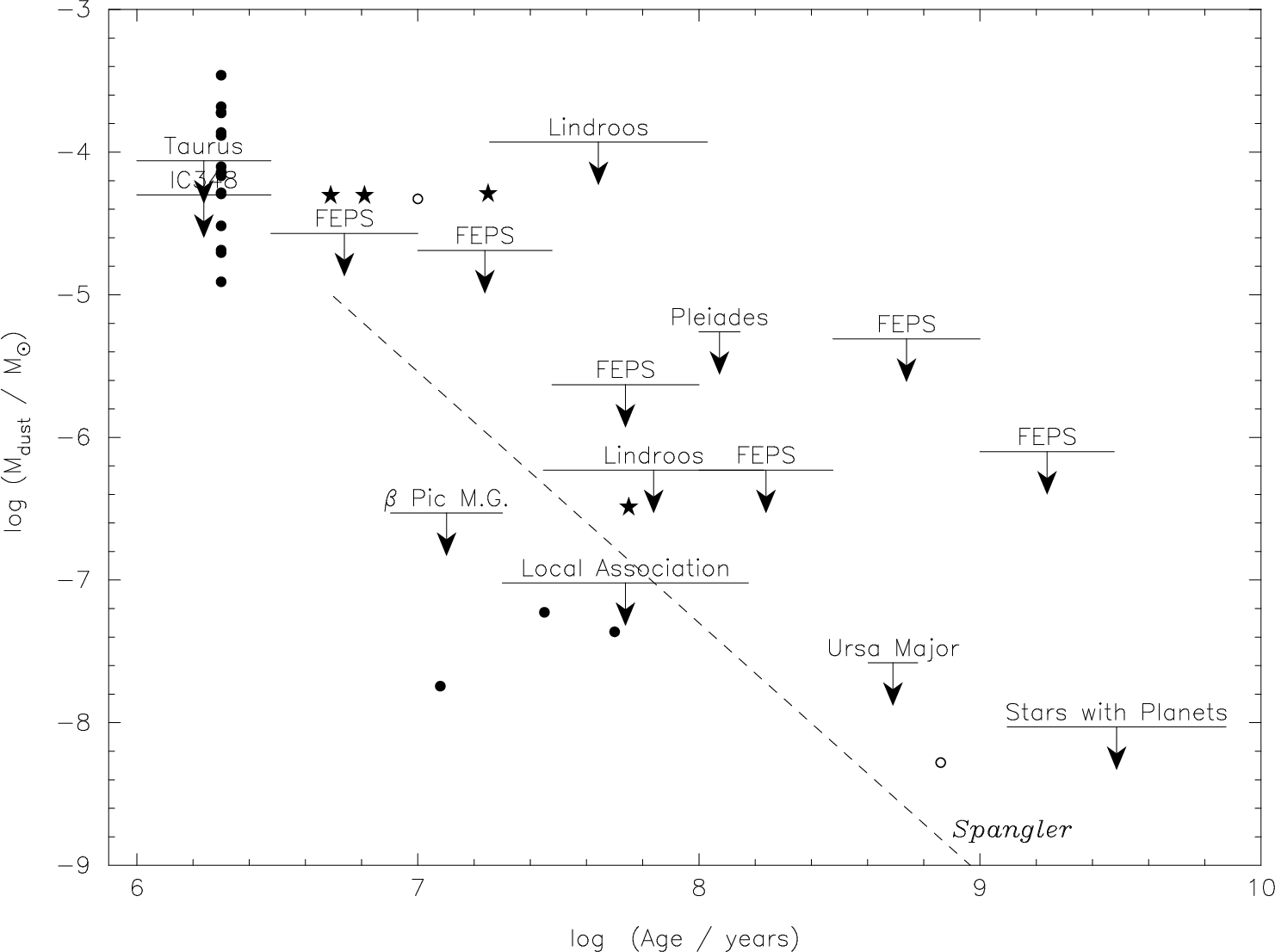}
\caption{
  \label{fig:mass}
  The mean circumstellar dust mass derived from millimeter or submillimeter 
  continuum observations (see Eq.~\ref{eq:mass}) as a function of age
  for stars with stellar masses between 0.5 and 2\msun. The stellar 
  samples shown include stars observed as part of the FEPS program as 
  presented in this paper, Taurus \citep{Beckwith90,Osterloh95,Duvert00}, 
  IC~348 \citep{Carp02}, Lindroos binary stars \citep{Jewitt94,Gahm94,Wyatt03},
  the $\beta$ Pic moving group and Local Association \citep{Liu04}, and stars 
  with known planets from radial velocity surveys \citep{Greaves04}. 
  Individual points indicate sources that have been detected at a signal to 
  noise ratio $\ge$ 3 in the FEPS sample (star symbols) and the other 
  stellar samples represented in this figure (solid circles). 
  The open circles show the location of TW~Hydra at 10~Myr \citep{Weintraub89} 
  and $\epsilon$~Eri \citep{Greaves98} at 730~Myr. The dotted line shows the
  mass-age relation derived by \citet{Spangler01} from {\it ISO} observations.
}
\end{figure}

The average amount of dust as a function of stellar age in the observed sample
can be examined by computing the mean dust mass and standard deviation of the 
mean in each age bin. If a star was observed at multiple wavelengths, the 
observation that provided the best sensitivity to dust mass for the adopted 
value of $\beta$ and $T_{\rm dust}$ was used in computing the mean values.
In principle, the mean dust mass for an ensemble of stars may yield a 
significant detection if many of the stars possess small quantities of dust 
which produces a small overall bias in the measured fluxes. In computing the 
averages, the stars were weighted uniformly since otherwise the derived values 
would be heavily weighted toward the few nearest stars that had the best 
sensitivity to dust mass, and therefore would not reflect the mass limits 
placed on the typical star in the sample. In each age bin, the mean dust mass 
was less than three times the standard deviation of the mean. 
Figure~\ref{fig:mass} shows the 3$\sigma$ upper limits to the mean dust masses 
for each age bin, which range from $5.9\times10^{-7}$\msun\ for the 
100-300~Myr stars to $2.5\times10^{-5}$~\msun\ for the 3-10~Myr age bin. The 
difference in the mass sensitivity limits as a function of age primarily 
reflects the fact that the younger stars are typically found at larger 
distances than the older stars in this sample.

\subsection{Comparison to Published Surveys}

To compare the results for the FEPS targets with other observations, we
compiled published millimeter and submillimeter continuum surveys of stars 
with stellar masses between 0.5 and 2.0\msun, which is the same mass range 
encompassed by the FEPS sample. These studies include the young stars in 
Taurus (40 stars total with the FEPS stellar mass range, with 16 millimeter 
continuum detections; Beckwith \etal~1990, Osterloh \& Beckwith~1995, Duvert 
\etal~2000), IC~348 (14 stars, with no detections; Carpenter~2002), 
the $\beta$ Pic moving group and Local Association (8 stars total, with 2 
detections; Liu \etal~2004), Lindroos binary systems (27 stars total, with 1 
detection; Jewitt~1994; Gahm \etal~1994; Wyatt, Dent, \& Greaves~2003), the 
Pleiades (12 stars with no detections; Zuckerman \& Becklin~1993), the Ursa 
Major moving group (12 stars with no detections; Zuckerman \& Becklin~1993), 
and stars with radial-velocity planets (8 stars total, with no detections; 
Greaves \etal~2004). The \citet{Wyatt03} observations of Lindroos binaries 
partially overlaps the sample observed by \citet{Jewitt94} and \citet{Gahm94},
and are significantly more sensitive. In presenting the results, we analyze 
the \citet{Wyatt03} data separately and only consider stars observed by
\citet{Jewitt94} and \citet{Gahm94} that were not observed at higher 
sensitivity. For the Taurus sample, stellar masses were estimated by first 
placing the stars in a Hertsprung-Russell diagram based on compiled photometry 
and spectral types, and then using the \citet{DM98} pre-main-sequence tracks 
to infer stellar masses (L. Hillenbrand, private communication). Similarly, 
for IC~348, we adopted the membership list, effective temperatures, and 
bolometric luminosities from \citet{Luhman03}, and also estimated stellar 
masses using the \citet{DM98} pre-main-sequence evolutionary tracks. In the 
remaining samples, we used the observed spectral type and estimated ages of 
the stellar group to estimate the stellar masses from evolutionary tracks.

Several continuum surveys for circumstellar dust around young stars
were not included in this analysis since there were few stars within the
desired stellar mass range or the continuum surveys were biased toward stars
with known infrared excesses. Continuum emission has been detected 
toward TW~Hydra \citep{Weintraub89} and reported for other members of the 
association \citep{Zuckerman01}, but an unbiased survey of the association
members has yet to be published. The extensive millimeter continuum surveys 
of $\rho$~Oph \citep{Andre94,Motte98,N98,Johnstone00}, NGC~2024 
\citep{Eisner03}, and Serpens \citep{Testi98} were omitted since these are 
heavily obscured regions and few stars have the necessary photometric and 
spectroscopic data to place the stars in the HR diagram and infer stellar 
masses. The continuum surveys of Lupus \citep{N97}, Chamaeleon~I 
\citep{Henning93} and MBM~12 \citep{Itoh03,Hogerheijde03} were not
included since few stars in these samples have stellar masses within the 
0.5-2.0\msun\ range. Continuum surveys of {\it IRAS}-detected debris disks 
\citep{Sylvester96,Holland98,Coulson98,Sylvester01,Holmes03,Sheret04} were 
excluded since they represent a biased sample of stars known to 
contain debris dust. 

The dust masses were re-computed from the observed fluxes in the surveys
described above using the assumptions for the dust properties adopted here. 
For those surveys that report the observed fluxes (as opposed to upper 
limits), the fluxes were averaged in a similar manner as was done for the 
FEPS targets to provide unbiased estimates of the mean dust mass. In each 
case, the mean was detected at a significance of less than 3$\sigma$. Some 
surveys provided only upper limits to the observed fluxes for stars with 
non-detections, and therefore the computed average disk mass is strictly an 
upper limit. Figure~\ref{fig:mass} shows the upper limits derived from 
published observations along with the upper limits from the FEPS targets. 
Since the mean dust mass is not detected at the 3$\sigma$ limit in any of 
the samples, these average values cannot be compared to 
test for evolution in the mean mass. Nonetheless, there is some suggestion of 
evolution in that the Taurus sample contains a number of sources with dust 
masses in excess of 10$^{-4}$\msun, but such massive disks are rare around 
stars with ages older than \about 10~Myr. These trends have been noted 
previously \citep{Zuckerman93,Jewitt94,Gahm94,Wyatt03,Liu04}.

\subsection{Temporal Evolution}

The evolution of dust masses was examined quantitatively using 
ASURV Rev 1.2 \citep{Lavalley92}, which implements the survival analysis
methods presented in \citet{Feigelson85}. Using the Gehan, logrank, and
Peto-Prentice tests, we computed the probability that the distribution of 
dust masses in any two stellar samples shown in Figure~\ref{fig:mass} could
have been drawn from the same parent population. Specifically, we used these
tests to determine which stellar samples have different dust mass distributions
than stars in Taurus, which may indicate the evolutionary time scale for 
cold dust in accretion disks. The probability that the Taurus sample and the 
3-10~Myr FEPS stars share the same parent population is between 0.098 and 
0.132 for the different tests. The corresponding probability, again in 
comparison to the Taurus sample, is between
a) 0.047 and 0.067 for the 10-30~Myr FEPS sample,
b) 0.003 and 0.005 for a combined 10-30~Myr sample that includes the FEPS 
sources, the $\beta$~Pic moving group, and Lindroos binaries that are within 
this age range, and 
c) $4.9\times10^{-6}$ and $1.3\times10^{-5}$ for the 30-100~Myr FEPS stars.

Before drawing conclusions from the comparisons between the different stellar
samples, we investigate the robustness of the results to various observational
uncertainties. The survival analysis routines do not incorporate 
uncertainties on a formal basis, and, therefore, we invoke an ad~hoc procedure
to address how uncertainties in the stellar ages and dust masses affect
the derived probabilities. We first consider the uncertainties in the 
stellar ages. The stellar ages for PDS~66, RX~J1842.9$-$3532 and 
RX~J1852.3$-$3700 are critical since these 3 stars have dust masses
comparable to that found in Taurus, and therefore the inclusion of these
stars in a particular age bin will increase the probability that the mass
distribution is similar to that found in Taurus. PDS~66 was placed in the
10-30~Myr age bin, but the derived age is between 7 and 17~Myr depending
on which pre-main-sequence evolutionary tracks are used \citep{Mamajek02},
and therefore the star may reasonably be placed in the 3-10~Myr age bin.
Given that RX~J1842.9$-$3532 and RX~J1852.3$-$3700 have estimated ages of
\about 10~Myr \citep{Neuhauser00}, it may be appropriate to assign these stars 
to the 10-30~Myr bin instead of 3-10~Myr. If we assign all three stars to the 
3-10~Myr bin, then the probability that the 3-10~Myr FEPS sample has the same 
dust mass distribution as Taurus increases from \about 0.11 to 0.19. If 
instead we assign all three stars to the 10-30~Myr bin, then the probability 
that the combined 10-30~Myr sample has the same mass distribution as Taurus 
increases from \about 0.003 to 0.017.

We assessed how uncertainties in the stellar distances and measured
fluxes (and consequently the derived dust masses) influence the survival 
analysis results by using a Monte Carlo simulation. We randomly assigned 
distances to each star using a gaussian random number generator that has a 
mean value centered on the nominal distance with a dispersion corresponding 
to the distance uncertainty. Similarly we varied the observed fluxes 
using both the observed flux measurement and calibration uncertainties. The 
dust masses were determined for these adjusted parameters, and the 
probabilities that the various samples 
are drawn from the same population were recomputed. We adopted a 10\% global 
uncertainty for the distance to Taurus \citep{Kenyon94}. For stars with 
Hipparcos parallaxes, we computed the distance uncertainty based on the 
parallax uncertainty, and for the remaining stars, we arbitrarily adopt an 
distance uncertainty of 20\%. In comparing the Taurus and 3-10~Myr FEPS 
sample, we found that the dispersion in the probabilities based on the 
measurement uncertainties is \about 0.04, and for Taurus and the combined 
10-30~Myr stellar samples, the dispersion is \about 0.002. Thus the largest 
source of uncertainty in comparing the stellar samples is assigning the stars 
to the appropriate age bin.

In summary, the above results show that the dust mass distribution for disks 
around 30-100~Myr stars is different from that in Taurus at high significance 
(\about 4.5$\sigma$). The mass distribution for the combined sample 10-30~Myr 
of stars is different from that of Taurus at the \about $2.5-3.4\sigma$ level, 
where the range reflects whether or not RX~J1842.9$-$3532 and 
RX~J1852.3$-$3700 are included in this age bin. Thus there is weak, but 
significant, evidence that the dust disks around 10-30~Myr solar-type stars 
are different than that found in Taurus. The differences in the dust mass 
distribution between Taurus and the 3-10~Myr FEPS sample is significant at the 
1.0-2.3$\sigma$ level, where the range reflects whether are not the 3 massive 
disks detected in this survey are included in this age bin. The current 
observational data are insufficient to determine if the circumstellar disk 
masses around solar-type stars, as traced by millimeter and submillimeter 
continuum observations, have evolved on time scales as short as 3-10~Myr.

\section{Discussion}
\label{discussion}

The differences in the dust mass distribution between the Taurus population 
and stars older than 10~Myr can be attributed to the relatively few massive 
disks found at older ages. The lack of old massive disks is unlikely to be an 
artifact of errors in the assumed dust temperatures. Any systematic variations 
in the dust temperature as a function of stellar age are such that the 
optically thick young disks are expected to be colder on average than the dust 
around older disks (cf. Beckwith \etal~1990 and Zuckerman \& Song~2004). Since 
the dust masses vary as $T_{dust}^{-1}$ in the Rayleigh-Jeans limit, the 
analysis presented here probably underestimates the differences in the dust 
mass distributions between the two samples. The apparent variations in dust 
mass distributions can likely be attributed then to a decrease in the amount of 
mass contained in small dust grains, and/or changes in the dust opacity due
to variations in the grain composition and the growth of grains into larger 
particles. These scenarios have interesting implications for the evolution of 
dust in disks, but the observations presented here cannot establish the 
dominant affect. 

We caution however that the evidence for evolution in the dust properties is 
relative to stars in the Taurus molecular cloud. Taurus is considered a 
representative region of ``isolated'', low mass star formation, while many of 
the 3-30~Myr stars surveyed here are members of the Scorpius-Centaurus
OB association. Environmental differences, as well as stellar age, then, may 
contribute to the differences in the mass distributions between Taurus and the 
10-30~Myr sample. \citet{Carp02} and \citet{Eisner03} in fact found that the 
dust mass distributions for stars in Taurus and the young clusters NGC~2024 
and IC~348 are different at the \about 2-3$\sigma$ level in that the cluster
regions contain few massive disks. These studies included a broader 
range of stellar masses than considered here, and the sample sizes of 
0.5-2\msun\ stars in these clusters is insufficient to identify any 
differences in the dust mass distributions over this narrower mass range.

The evolution of disk properties for ages \aboutless 30~Myr inferred from 
millimeter continuum observations are qualitatively consistent with that found 
from near- and mid-infrared observations. Nearly all low-mass stars with ages 
of \about 1~Myr have near-infrared excesses characteristic of circumstellar 
disks \citep{Strom89,Haisch01}. Only 1\% of the solar-type stars in Lower 
Centaurus-Crux and Upper Centaurus-Lupus possess such inner disks, suggesting
that the inner disk dissipates on time scales less than 15~Myr
\citep{Mamajek02}. Observations at 10\micron\ (Mamajek \etal~2004 and 
references therein) and 60\micron\ \citep{Meyer00} also show that a lower
fraction of 10-30~Myr stars exhibit an excess at these wavelengths relative to 
1~Myr stars. Thus observations at wavelengths from 2\micron\ to 3~mm all 
point to a decrease in the number of stars with disks between \about 1~Myr and
\about 10-30~Myr. However, the nature of the decline in the disk frequency at
ages of 3-10~Myr remains ambiguous. \citet{Haisch01} suggest that the
inner disk frequency diminishes to near zero by an age of 6~Myr, while 
\citet{Lyo03} suggest that \about 60\% of the stars in the 5-9~Myr 
$\eta$~Cha cluster contain an inner disk. Available millimeter observations 
are insufficient to distinguish differences in the dust mass distributions 
around 3-10~Myr stars relative to Taurus. 

The temporal evolution in the mass of warm dust around solar-type stars has 
been investigated by \citet{Spangler01} and \citet{Habing01} using {\it ISO} 
60\micron, 90\micron, 100\micron, and 170\micron\ observations. Both surveys 
find a decrease in the dust luminosity with stellar age. \citet{Habing01} 
though suggests a dramatic decrease in the number of debris disks 
for stars older than \about 400~Myr, while \citet{Spangler01} suggested that 
the mean fractional dust luminosity, $f_d$, declines steadily with stellar age 
as $f_d \propto t^{-1.76}$. Figure~\ref{fig:mass} shows the temporal evolution 
of dust mass derived from the {\it ISO} observations using the approximate 
relation between $f_d$ and dust mass from \citet{Silverstone00}. The dust
mass upper limits from the millimeter-wavelength observations are generally
above the \citet{Spangler01} relation and therefore do not resolve the 
discrepancy between the \citet{Habing01} and \citet{Spangler01} results. 
The lack of millimeter-wavelength continuum detections does suggest that 
there are not massive reservoirs of cold dust (\aboutless 20~K; see 
Fig.~\ref{fig:tdust}) that has been missed by {\it IRAS} and {\it ISO} 
observations. 

\section{Summary}
\label{summary}

We present submillimeter (CSO 350\micron) and millimeter (SEST 1.2~mm, OVRO 
3~mm) photometry 
for 125 stars that will be observed by the FEPS {\it Spitzer} Legacy Program. 
These stars have stellar masses between 0.5 and 2\msun\ and stellar ages 
between \about 3~Myr and 1~Gyr, and are used to investigate the evolution of
cold circumstellar dust around solar-type stars. Four sources in this survey 
were detected in the millimeter continuum: RX~J1842.9$-$3532, 
RX~J1852.3$-$3700, and PDS~66 with SEST, and HD~107146 with OVRO. 
RX~J1842.9$-$3532 and RX~J1852.3$-$3700 are located in projection near the 
RCrA molecular cloud with estimated ages of \about 10~Myr \citep{Neuhauser00}.
PDS~66 is a kinematic member of the \about 20~Myr old Lower Centaurus-Crux 
subgroup of the Scorpius-Centaurus OB association and is probably a rare 
example of a old classical T Tauri star surrounded by an accretion disk 
\citep{Mamajek02}. HD~107146 is a young (80-200~Myr) debris disk system that 
was recently detected in the submillimeter continuum \citep{Williams04}. 

The SEST detections of RX~J1842.9$-$3532, RX~J1852.3$-$3700, and PDS~66 are 
unresolved at a FWHM resolution of 24$''$, and the observed fluxes imply dust 
masses of \about 5$\times10^{-5}$\msun\ around each star, assuming a dust 
temperature of 40~K, a mass absorption coefficient of 2~cm$^2$g$^{-1}$ at 
$\lambda$1.3~mm, and $\beta=1$ \citep{Beckwith90}. Since these three 
stars have observational characteristics similar to that of classical T Tauri 
stars, the continuum emission likely originates from a circumstellar 
accretion disk. The OVRO observations of HD~107146 resolve the continuum 
emission with a gaussian-fit FWHM size of 
$(6.5''\pm1.4'') \times (4.2''\pm1.3'')$, or 185~AU$\times$120~AU, that is 
centered on the stellar position within the astrometric uncertainties.

To investigate the evolution of cold circumstellar dust around 0.5-2.0\msun\
stars, our results are compared with published continuum observations of 
stars in Taurus \citep{Beckwith90,Osterloh95,Duvert00}, the $\beta$~Pic moving 
group and Local Association \citep{Liu04}, and Lindroos binaries 
\citep{Jewitt94,Gahm94,Wyatt03}. The stars were grouped into age bins that 
span a factor of 3, and are compared with the dust masses inferred around 
solar-type stars in Taurus to investigate the evolutionary time scales for 
dust mass in circumstellar disks. Using ``survival 
analysis'' techniques \citep{Feigelson85}, the mass distribution of disks
in the Taurus molecular cloud is distinguished from that in the 10-30~Myr 
stellar sample at a significance level of 2.5-3.4$\sigma$, where the
range in significance reflects the uncertainty in determining the 
age of stars that have continuum detections. The difference in the dust
mass distributions between Taurus and the 3-10~Myr stars is significant
at the 1.0-2.3$\sigma$ confidence level. These results suggest that 
significant evolution has occurred in the circumstellar dust properties 
around solar-type stars by ages of 10-30~Myr, either by a decrease
in the mass of small dust grains with time and/or changes in the dust opacity.
These time scales are consistent with that inferred for the inner disk as 
traced by 2-60\micron\ infrared emission. Additional observations are needed 
to establish if evolution occurs on even shorter time scales.

\acknowledgements

We gratefully acknowledge the FEPS team, especially Erik Mamajek, John 
Stauffer, and Lynne Hillenbrand, for their efforts in defining the FEPS sample.
We would also like to thank Michael Meyer for detailed comments on the paper, 
and Darren Dowell for his assistance with the CSO observations and data 
reduction. JMC acknowledges support from the Long Term Space Astrophysics Grant 
NAG5-8217, the {\it Spitzer} Legacy Science Program through an award issued by 
JPL/CIT under NASA contract 1407, and the Owens Valley Radio Observatory, 
which is supported by the National Science Foundation through grant 
AST-9981546. S.~Wolf is supported through the Emmy Noether grant WO 857/2-1 of 
the German Research Foundation, the NASA grant NAG5-11645, and the 
{\it Spitzer} Legacy Science Program NASA contract 1407 to JPL/CIT.
The Caltech Submillimeter Observatory is supported by NSF grant AST 02-29008.
This publication makes use of data products from the Two Micron All Sky 
Survey, which is a joint project of the University of Massachusetts and the 
Infrared Processing and Analysis Center, funded by the National Aeronautics 
and Space Administration and the National Science Foundation. This research 
has made use of the SIMBAD database, operated at CDS, Strasbourg, France.

\clearpage

\clearpage

\clearpage

\clearpage

\clearpage

\clearpage

\clearpage

\clearpage

\clearpage

\input{Carpenter.tab1}
\end{document}

%% file: Carpenter.tab1.tex
%

\tablenum{1}

\begin{deluxetable}{lrrrrr}
\tablewidth{470pt}
\tablecaption{Observed Sources\label{tbl:sources}}
\tablehead{
  \colhead{Name}       &
  \colhead{Distance}   &
  \colhead{S$_\nu$(350$\mu$m)} &
  \colhead{S$_\nu$(1.2 mm)} &
  \colhead{S$_\nu$(2.7 mm)} &
  \colhead{S$_\nu$(3.0 mm)} \\
  \cline{3-6}\\
  \colhead{}           & 
  \colhead{(pc)}       & 
  \multicolumn{4}{c}{(mJy)}
}
\startdata
\multicolumn{6}{c}{\it 6.5 $\le$ log(age/years)  $<$ 7.0 }\\
HD 104467 & 110 & \nodata  & $-15\pm21$ & \nodata  & \nodata  \\
HD 142361 & 81 & \nodata  & \nodata  & \nodata  & $1.59\pm1.11$ \\
HD 284135 & 140 & $<135$ & \nodata  & \nodata  & \nodata  \\
HD 285372 & 140 & $<134$ & \nodata  & \nodata  & \nodata  \\
RX J1111.7-7620\tablenotemark{*} & 160 & \nodata  & $17\pm10$ & \nodata  & \nodata  \\
RX J1541.1-2656 & 145 & \nodata  & \nodata  & \nodata  & $-1.07\pm1.44$ \\
RX J1842.9-3532\tablenotemark{*} & 130 & \nodata  & $88\pm6$ & \nodata  & \nodata  \\
RX J1852.3-3700\tablenotemark{*} & 130 & \nodata  & $88\pm8$ & \nodata  & \nodata  \\
RX J1917.4-3756 & 130 & \nodata  & $3\pm18$ & \nodata  & \nodata  \\
ScoPMS 27 & 145 & \nodata  & $-1\pm6$ & $-3.82\pm2.33$ & \nodata  \\
ScoPMS 52 & 145 & \nodata  & \nodata  & \nodata  & $0.09\pm0.71$ \\
ScoPMS 214 & 145 & \nodata  & \nodata  & \nodata  & $0.58\pm1.21$ \\
$[$PZ99$]$ J155847.8-175800 & 160 & \nodata  & \nodata  & \nodata  & $0.74\pm1.59$ \\
$[$PZ99$]$ J161329.3-231106 & 145 & \nodata  & $-31\pm20$ & \nodata  & $-0.94\pm0.84$ \\
\multicolumn{6}{c}{}\\
\multicolumn{6}{c}{\it 7.0 $\le$ log(age/years)  $<$ 7.5 }\\
AO Men & 38 & \nodata  & $26\pm30$ & \nodata  & \nodata  \\
HD 107441 & 97 & \nodata  & $-16\pm17$ & \nodata  & \nodata  \\
HD 139498 & 124 & \nodata  & \nodata  & \nodata  & $1.37\pm1.46$ \\
HD 199143 & 48 & \nodata  & $-8\pm8$ & \nodata  & \nodata  \\
MML 1 & 83 & \nodata  & $<60$ & \nodata  & \nodata  \\
MML 8 & 96 & \nodata  & $-1\pm20$ & \nodata  & \nodata  \\
MML 9 & 95 & \nodata  & $-22\pm16$ & \nodata  & \nodata  \\
MML 17 & 110 & \nodata  & $-8\pm18$ & \nodata  & \nodata  \\
MML 18 & 100 & \nodata  & $-16\pm18$ & \nodata  & \nodata  \\
PDS 66\tablenotemark{*} & 86 & \nodata  & $207\pm11$ & \nodata  & \nodata  \\
V343 Nor & 40 & \nodata  & $-1\pm7$ & \nodata  & \nodata  \\
\multicolumn{6}{c}{}\\
\multicolumn{6}{c}{\it 7.5 $\le$ log(age/years)  $<$ 8.0 }\\
1RXS J043243.2-152003 & 140 & \nodata  & $-13\pm13$ & \nodata  & \nodata  \\
1RXS J053650.0+133756 & 100 & $<150$ & \nodata  & \nodata  & \nodata  \\
B102 & 150 & \nodata  & $14\pm19$ & \nodata  & \nodata  \\
HD 105 & 40 & \nodata  & $11\pm27$ & \nodata  & \nodata  \\
HD 377 & 40 & \nodata  & \nodata  & $0.32\pm0.80$ & $0.79\pm0.61$ \\
HD 984 & 46 & \nodata  & $13\pm30$ & \nodata  & $0.35\pm0.58$ \\
HD 12039 & 42 & \nodata  & $-2\pm24$ & \nodata  & $-0.38\pm0.63$ \\
HD 15526 & 100 & \nodata  & $-12\pm10$ & \nodata  & \nodata  \\
HD 17925 & 10 & \nodata  & $13\pm34$ & \nodata  & $-0.81\pm0.55$ \\
HD 25457\tablenotemark{*} & 19 & \nodata  & $-8\pm14$ & $-1.51\pm1.33$ & $0.37\pm0.62$ \\
HD 35850 & 27 & \nodata  & $-8\pm36$ & $0.84\pm1.59$ & $0.07\pm0.72$ \\
HD 37484\tablenotemark{*} & 60 & \nodata  & $7\pm15$ & \nodata  & \nodata  \\
HD 47875 & 70 & \nodata  & $10\pm16$ & \nodata  & \nodata  \\
HD 60737 & 38 & \nodata  & \nodata  & \nodata  & $-1.70\pm1.01$ \\
HD 61005\tablenotemark{*} & 35 & \nodata  & $31\pm34$ & \nodata  & \nodata  \\
HD 70516 & 37 & \nodata  & \nodata  & \nodata  & $0.06\pm0.97$ \\
HD 77407 & 30 & \nodata  & \nodata  & \nodata  & $-0.10\pm0.62$ \\
HD 86356 & 110 & \nodata  & $-17\pm17$ & \nodata  & \nodata  \\
HD 104860 & 48 & \nodata  & \nodata  & \nodata  & $1.35\pm0.67$ \\
HD 107146\tablenotemark{*} & 29 & \nodata  & \nodata  & \nodata  & $1.42\pm0.23$ \\
HD 112196 & 34 & \nodata  & \nodata  & \nodata  & $-0.17\pm0.53$ \\
HD 129333 & 34 & \nodata  & \nodata  & \nodata  & $-0.18\pm0.60$ \\
HD 134319 & 44 & \nodata  & \nodata  & $1.06\pm0.80$ & $0.17\pm0.23$ \\
HD 135363 & 29 & \nodata  & \nodata  & \nodata  & $1.27\pm0.77$ \\
HD 151798 & 41 & \nodata  & \nodata  & \nodata  & $-0.15\pm0.58$ \\
HD 202917 & 46 & \nodata  & $-1\pm37$ & \nodata  & \nodata  \\
HD 216803 & 8 & \nodata  & $-31\pm27$ & \nodata  & \nodata  \\
HD 217343 & 32 & \nodata  & $32\pm29$ & \nodata  & $-0.31\pm0.66$ \\
HD 279788 & 100 & $<177$ & \nodata  & \nodata  & \nodata  \\
R3 & 150 & \nodata  & $12\pm16$ & \nodata  & \nodata  \\
R45 & 150 & \nodata  & $4\pm15$ & \nodata  & \nodata  \\
R83 & 150 & \nodata  & $11\pm18$ & \nodata  & \nodata  \\
RE J0137+18A & 61 & $<131$ & \nodata  & \nodata  & \nodata  \\
RX J0331.1+0713 & 100 & $<137$ & \nodata  & \nodata  & \nodata  \\
RX J1140.3-8321 & 110 & \nodata  & $-10\pm19$ & \nodata  & \nodata  \\
RX J1844.3-3541 & 130 & \nodata  & $12\pm20$ & \nodata  & \nodata  \\
SAO 150676 & 65 & \nodata  & $6\pm12$ & \nodata  & $-0.11\pm0.71$ \\
V383 Lac & 50 & \nodata  & \nodata  & \nodata  & $-0.89\pm0.62$ \\
W79 & 150 & \nodata  & $-8\pm18$ & \nodata  & \nodata  \\
\multicolumn{6}{c}{}\\
\multicolumn{6}{c}{\it 8.0 $\le$ log(age/years)  $<$ 8.5 }\\
2RE J0255+474 & 50 & \nodata  & \nodata  & $-0.91\pm1.53$ & $-0.35\pm0.30$ \\
BPM 87617 & 50 & \nodata  & \nodata  & \nodata  & $-0.92\pm0.53$ \\
HD 691 & 34 & \nodata  & \nodata  & \nodata  & $-0.02\pm0.77$ \\
HD 8907\tablenotemark{*} & 34 & \nodata  & \nodata  & $-0.05\pm0.45$ & $-0.18\pm0.37$ \\
HD 19668 & 40 & \nodata  & $-18\pm10$ & \nodata  & $0.38\pm0.64$ \\
HD 25300 & 50 & \nodata  & $-7\pm13$ & \nodata  & \nodata  \\
HD 38207\tablenotemark{*} & 100 & \nodata  & $-3\pm12$ & \nodata  & \nodata  \\
HD 38949 & 43 & \nodata  & $-10\pm12$ & \nodata  & \nodata  \\
HD 41700 & 27 & \nodata  & $9\pm12$ & \nodata  & \nodata  \\
HD 70573 & 46 & \nodata  & \nodata  & \nodata  & $0.02\pm0.70$ \\
HD 72905 & 14 & \nodata  & \nodata  & \nodata  & $0.59\pm0.59$ \\
HD 75393 & 42 & \nodata  & $11\pm32$ & \nodata  & \nodata  \\
HD 88201 & 41 & \nodata  & $17\pm14$ & \nodata  & \nodata  \\
HD 90905 & 32 & \nodata  & \nodata  & \nodata  & $-0.29\pm0.56$ \\
HD 91782 & 56 & \nodata  & \nodata  & \nodata  & $-0.64\pm0.68$ \\
HD 91962 & 37 & \nodata  & $-24\pm66$ & \nodata  & \nodata  \\
HD 92855 & 36 & \nodata  & \nodata  & \nodata  & $0.97\pm0.47$ \\
HD 95188 & 36 & \nodata  & \nodata  & \nodata  & $-0.22\pm0.57$ \\
HD 101472 & 39 & \nodata  & $-1\pm16$ & \nodata  & $-0.66\pm0.87$ \\
HD 104576 & 49 & \nodata  & $-7\pm17$ & \nodata  & $-0.20\pm0.75$ \\
HD 106772 & 110 & \nodata  & $11\pm25$ & \nodata  & \nodata  \\
HD 108799 & 25 & \nodata  & \nodata  & \nodata  & $-0.92\pm1.17$ \\
HD 108944 & 44 & \nodata  & \nodata  & \nodata  & $-0.35\pm0.52$ \\
HD 132173 & 49 & \nodata  & \nodata  & \nodata  & $0.39\pm1.26$ \\
HD 139813\tablenotemark{*} & 22 & \nodata  & \nodata  & \nodata  & $-0.05\pm0.31$ \\
HD 152555 & 48 & \nodata  & \nodata  & \nodata  & $-0.02\pm0.44$ \\
HD 172649 & 47 & \nodata  & \nodata  & \nodata  & $0.65\pm0.48$ \\
HD 191089\tablenotemark{*} & 54 & \nodata  & $6\pm12$ & \nodata  & $-0.18\pm0.61$ \\
HD 199019 & 35 & \nodata  & \nodata  & \nodata  & $-0.07\pm0.62$ \\
HD 203030 & 41 & \nodata  & \nodata  & \nodata  & $0.65\pm0.44$ \\
HD 209253\tablenotemark{*} & 30 & \nodata  & $12\pm19$ & \nodata  & \nodata  \\
HD 224873 & 49 & \nodata  & \nodata  & \nodata  & $-0.38\pm0.49$ \\
HIP 6276 & 35 & \nodata  & $7\pm8$ & \nodata  & $1.07\pm0.67$ \\
QT And & 50 & \nodata  & \nodata  & \nodata  & $0.47\pm0.59$ \\
RE J0723+20 & 24 & \nodata  & \nodata  & \nodata  & $-0.99\pm0.73$ \\
RX J0849.2-7735 & 90 & \nodata  & $2\pm13$ & \nodata  & \nodata  \\
RX J0850.1-7554 & 90 & \nodata  & $-19\pm18$ & \nodata  & \nodata  \\
SAO 178272 & 58 & \nodata  & $5\pm45$ & \nodata  & \nodata  \\
vB 142 & 51 & \nodata  & \nodata  & \nodata  & $-0.16\pm0.78$ \\
\multicolumn{6}{c}{}\\
\multicolumn{6}{c}{\it 8.5 $\le$ log(age/years)  $<$ 9.0 }\\
HD 7661 & 27 & \nodata  & $-4\pm6$ & \nodata  & \nodata  \\
HD 21411 & 31 & \nodata  & $17\pm29$ & \nodata  & \nodata  \\
HD 209779 & 36 & \nodata  & $-53\pm42$ & \nodata  & \nodata  \\
HIP 59154 & 43 & \nodata  & $36\pm21$ & \nodata  & \nodata  \\
RX J0917.2-7744 & 90 & \nodata  & $-6\pm17$ & \nodata  & \nodata  \\
RX J1203.7-8129 & 110 & \nodata  & $-6\pm22$ & \nodata  & \nodata  \\
RX J1209.8-7344 & 110 & \nodata  & $-25\pm19$ & \nodata  & \nodata  \\
RX J1220.6-7539 & 110 & \nodata  & $-24\pm22$ & \nodata  & \nodata  \\
\multicolumn{6}{c}{}\\
\multicolumn{6}{c}{\it 9.0 $\le$ log(age/years)  $<$ 9.5 }\\
HD 6434 & 40 & \nodata  & $0\pm10$ & \nodata  & \nodata  \\
HD 27466 & 36 & \nodata  & $10\pm13$ & \nodata  & \nodata  \\
HD 29231 & 28 & \nodata  & $-9\pm16$ & \nodata  & \nodata  \\
HD 31143 & 32 & \nodata  & $0\pm15$ & \nodata  & \nodata  \\
HD 37962 & 37 & \nodata  & $-8\pm12$ & \nodata  & \nodata  \\
HD 69076 & 34 & \nodata  & $3\pm37$ & \nodata  & \nodata  \\
HD 88742 & 23 & \nodata  & $-21\pm13$ & \nodata  & \nodata  \\
HD 92788 & 32 & \nodata  & $5\pm15$ & \nodata  & \nodata  \\
HD 98553 & 34 & \nodata  & $27\pm41$ & \nodata  & \nodata  \\
HD 101959 & 32 & \nodata  & $2\pm19$ & \nodata  & \nodata  \\
HD 102071 & 30 & \nodata  & $76\pm40$ & \nodata  & \nodata  \\
HD 179949 & 27 & \nodata  & $1\pm10$ & \nodata  & \nodata  \\
HD 183216 & 35 & \nodata  & $-14\pm15$ & \nodata  & \nodata  \\
HD 193017 & 37 & \nodata  & $-11\pm17$ & \nodata  & \nodata  \\
HD 201989 & 30 & \nodata  & $31\pm27$ & \nodata  & \nodata  \\
HD 205905 & 26 & \nodata  & $37\pm26$ & \nodata  & \nodata  \\
\multicolumn{6}{c}{}\\
\multicolumn{6}{c}{\it Sources dropped from the FEPS program}\\
B119 & 150 & \nodata  & $-10\pm17$ & \nodata  & \nodata  \\
B140 & 150 & \nodata  & $-8\pm16$ & \nodata  & \nodata  \\
HD 19632 & 30 & \nodata  & $19\pm15$ & \nodata  & \nodata  \\
HD 22879 & 24 & \nodata  & $1\pm15$ & \nodata  & \nodata  \\
HD 49197 & 45 & \nodata  & \nodata  & \nodata  & $0.26\pm0.35$ \\
HD 88638 & 38 & \nodata  & \nodata  & \nodata  & $0.23\pm0.52$ \\
HD 99565 & 35 & \nodata  & $26\pm45$ & \nodata  & \nodata  \\
HD 166435 & 25 & \nodata  & \nodata  & \nodata  & $0.86\pm0.53$ \\
HD 221275 & 35 & \nodata  & $-33\pm73$ & \nodata  & \nodata  \\
RX J1017.9-7431 & 110 & \nodata  & $-8\pm17$ & \nodata  & \nodata  \\
RX J1307.3-7602 & 110 & \nodata  & $-5\pm22$ & \nodata  & \nodata  \\
\enddata
\tablenotetext{*}{Contains an infrared excess in one or more {\it IRAS} bands.}
\end{deluxetable}